\documentclass{aa}
\usepackage{graphicx}

\begin{document}

\title{Spherically symmetric relativistic MHD simulations\\
of pulsar wind nebulae in supernova remnants}

\author{
N. Bucciantini \inst{1,2}, J.~M. Blondin\inst{2}, L. Del Zanna \inst{1}, E. Amato \inst{3}}

\offprints{N.Bucciantini\\
 e-mail: niccolo@arcetri.astro.it}

\institute{Dip. di Astronomia e Scienza dello Spazio,
  Universit\`a di Firenze, Largo E.Fermi 2, I-50125 Firenze, Italy
\and
Department of Physics, North Carolina State University,
Raleigh, NC 27695, USA
\and
INAF, Osservatorio Astrofisico di Arcetri,
 Largo E.Fermi 5, I-50125 Firenze, Italy
}
\authorrunning{N.Bucciantini et al.}
\titlerunning{Spherically symmetric RMHD simulations of pulsar wind nebulae in SNRs}

\date{Received 24 February 2003/ Accepted 21 March 2003}
\abstract{Pulsars, formed during  supernova explosions, are known to be sources of relativistic magnetized winds whose interaction with the expanding supernova remnants (SNRs) gives rise to a pulsar wind nebula (PWN). We present spherically symmetric  relativistic magnetohydrodynamics (RMHD) simulations of the interaction of a pulsar wind with the surrounding SNR, both in particle and magnetically dominated regimes. As shown by previous simulations, the evolution can be divided in three phases: free expansion, a transient phase characterized by the compression and reverberation of the reverse shock, and a final Sedov expansion. The evolution of the contact discontinuity between the PWN and the SNR (and consequently of the SNR itself) is almost independent of the magnetization of the nebula as long as the total (magnetic plus particle) energy is the same. However, a different behaviour of the PWN internal structure is observable during the compression-reverberation phase, depending on the degree of magnetization. The simulations were performed using the third order conservative scheme by Del Zanna et al. (2003). 

}

\maketitle

\section{Introduction}

Supernova remnants (SNRs) are nebulae originated by the explosions of massive stars, known as Supernova events (SN), which tipically release an amount of energy $\sim 10^{53}$ erg. Most of this energy is carried away by neutrinos produced during the core-collapse phase and the formation of a degenerate stellar remnant (neutron star). The remaining energy (about 1\%) gives rise to a blast wave that sweeps up the outer layers of the star and produces a strong shock propagating in the surrounding medium. The details of such an expansion depend on a number of different parameters: the ejected mass and energy, the nature and density distribution of the ambient medium (\cite{dwarkadas98}, \cite{featherstone2001}, \cite{blondin96}), the efficiency of radiative losses (\cite{chevalier95}), anisotropies in the supernova explosion (\cite{wang2002}, \cite{chevalier89}), as well as neutron star spin-down power and proper motion space velocity (\cite{chatterjee02}, \cite{frail94}, \cite{strom87}). Hence, in principle one should expect to find a large variety of different structures among PWN-SNR systems.

If the stellar remnant is a rapidly spinning magnetized neutron star (pulsar), then a late energy input is supplied to the nebular remnant in the form of a wind composed of relativistic particles (mainly electron-positron pairs, with possibly a minority of ions (\cite{amato03} and references within)) and a toroidal magnetic field. Most of the pulsar rotational energy goes into the launching of this believed to be highly relativistic wind, with a bulk Lorentz factor tipically estimated to be in the range  $10^{4}-10^{7}$ (\cite{rees67}, \cite{michel01}). The detailed magnetospheric physics that is at the origin of such an outflow is still poorly understood, but one point on which all current theories agree is that the wind should be magnetically dominated at a distance from the pulsar corresponding to the so called light cylinder radius, $R_{\rm LC}=c/\Omega$, with $\Omega$ the pulsar spin frequency. At larger distances a bubble of relativistically hot magnetized plasma is then created. This bubble (often called ``plerion'') shines in a very large range of frequencies, from radio wavelengths up to X-rays and even $\gamma$-rays, due to the synchrotron and Inverse Compton emission of the relativistic particles. The global properties of the pulsar wind nebula (PWN) can be modeled in terms of injection parameters and evolutionary effects (\cite{pacini}, \cite{reynolds84}). However the energy released by the pulsar during its entire life-time ($\sim 10^{49}$ erg in the case of the Crab Pulsar) is much less than that driving the expansion and evolution of the SNR. Therefore we expect the PWN to have limited effects on the global dynamics of the SNR.

The evolution of the PWN-SNR can be divided into various phases with
different observational features (\cite{woltjer72}, \cite{cioffi90},
\cite{reynolds84}, \cite{chevalier98}), which result from the
interaction of the SNR structures with the pulsar wind bubble. The SN
explosion leads to the formation of a SNR in which three main
discontinuities can be found: a forward shock that propagates in the
ISM, a reverse shock that propagates inside the free expanding
ejecta, and  a contact discontinuity separating the shocked ISM from
the shocked SN material. Initially the overall structure expands and
also the reverse shock moves away from the origin of the explosion,
but, as the amount of swept ISM grows, the speed of the contact
discontinuity decreases and, finally, the reverse shock starts moving
back to the origin, where it collapses on a time scale of
order $10^{3}-10^{4}$ years. Initially (up to about $10^{3}$ years
after the SN explosion, before the reverse shock in the ejecta starts
moving back), the pulsar spin-down luminosity  is nearly constant and
provides a steady energy injection in the PWN.  The PWN expands as $t^{6/5}$ in the freely expanding supernova ejecta, in the case where the SN has a constant density profile (we will refer to this phase as free expansion phase due to the fact that the PWN is inside the freely expanding ejecta). The second phase (often referred as Sedov-Taylor phase) results when the PWN contact discontinuity reaches the SNR reverse shock in the ejecta. At this point the PWN expansion stops and the SNR reverse shock starts compressing the PWN toward the pulsar (\cite{mckee74}, \cite{cioffi88}). The main observational signature of this process is an increase in radio luminosity associated to the magnetic field enhancement and particle re-energization due to compression. In the absence of a PWN the SNR reverse shock would collapse to the center, but when a plerion exists the presence of an hot bubble prevents the collapse from happening and after a transient phase, characterized by oscillations of the contact discontinuity, the nebula enters  a phase of slow expansion (Sedov expansion phase (\cite{sedov59})), and finally dissipates in the interstellar medium (ISM).

Hydrodynamical simulations aimed at investigating the free expansion and Sedov-Taylor phases were recently carried out both in one (\cite{swaluw01} (SW)) and two (\cite{blondin01} (BCF), \cite{jun98}) dimensions. These studies were intended to provide some insights into the details of the system evolution and its stability properties (Rayleigh-Taylor instability may occur both in the free expansion (\cite{jun98}), and in the Sedov-Taylor (BCF) stages as a consequence of ejecta acceleration and compression by the reverse shock). However, all the previous investigations dealt with non-relativistic hydrodynamic regimes. Our aim is to extend such simulations to the more realistic relativistic magnetohydrodynamical (RMHD hereafter) regime, in order to evaluate if the hydrodynamical (HD hereafter) approximation is good enough and what differences one may expect. In this paper, we present one-dimensional RMHD simulations of the first two evolutionary stages in spherical symmetry, thus extending the HD study by SW. Two-dimensional simulations of the free-expansion phase will be presented in a forthcoming paper. We will show that no significant differences in the global evolution arise between the HD and RMHD cases, both in the free expansion phase and during the reverse shock reverberation, even if different structures may form inside the PWN. We expect major differences to arise in the multidimensional case were the toroidal magnetic field may play an essential role in the development and growth of instabilities, eventually removing degrees of freedom of the system.

\section{Numerical Simulation}

All the simulations have been performed by using the newly developed scheme by Del Zanna et al. (\cite{delzanna02}, \cite{delzanna03}). We refer the reader to the cited papers for a detailed description of the code, the equations and algorithms employed. This is a high resolution conservative (shock-capturing) code for 3D-RMHD based on accurate third order reconstruction ENO-type algorithms and on an approximate Riemann solver flux formula (HLL) which does not make use of time-consuming characteristic decomposition. The code used here has been modified by adding a new equation for a tracer that allows us to use different adiabatic coefficients for the pulsar wind material and for the SNR-ISM medium, namely $\gamma=4/3$ and $\gamma=5/3$ respectively, following the same approach as in Bucciantini (2002), but modified for the HLL solver actually used by the code. The use of two different adiabatic coefficients on a contact discontinuity with a very large density jump (density may change by factors over ~$10^{6}-10^{7}$, see Fig.~2) leads to the formation of waves that tend to propagate back into the PWN (\cite{shyue98}, \cite{karni94}); however such waves are well subsonic (the ratio between velocity fluctuations and the sound speed is about 0.07 in the HD case, and that between kinetic and thermal energy is less than $10^{-3}$) so that the behaviour of the PWN is not very much affected.

\subsection{Choice of initial parameter and injection conditions}
The simulations were performed on a 1024-cell radial grid, corresponding to a physical domain extending from the origin  to 30 Ly. The evolution of the system is followed for 30,000 years. We set continuous conditions at the outer boundary (zeroth order extrapolation) and reflection at the origin. Density, momentum-energy and magnetic field from the pulsar are injected in the first cell. No radiation cooling is assumed. 

Initial conditions are similar to those adopted by SW: total energy in
the ejecta $E_{tot}=10^{51}$ erg, mass in the ejecta
$M_{ej}=3M_{\odot}$, ISM mass density $10^{-24}$g/cm$^{3}$, 
  chosen to match the partameters for the Crab Nebula. The supernova ejecta are set at the initial time in the first computational cells: we adopt here a spatially constant density ($\rho_{o}$) profile and a velocity proportional to the distance from the origin such that:
\begin{eqnarray}
M_{ej}=\int_{0}^{R_{o}}4\pi\rho_{o}r^{2}dr \\
E_{tot}\simeq\int_{0}^{R_{o}}\frac{4\pi}{2}\rho_{o}v^{2}r^{2}dr
\end{eqnarray}
where $R_{o}$ is the initial radius of the ejecta (we have set $R_{o}=1.2$ Ly for a good resolution, but simulations give nearly the same results also for smaller values). While in the work by  SW the supernova energy is initially set as enhanced thermal pressure, here we have decided to set it as kinetic energy for numerical convenience, in order to reduce the initial diffusion of the contact discontinuity and thus to obtain  a sharper density profile (this problem is common to all {\em central}-type schemes that avoid spectral decomposition).

The pulsar wind is created by injecting mass, momentum-energy, and a purely toroidal magnetic field in the first computational cell, with a total luminosity that depends on time in order to include the spin-down process:
\begin{equation}
L=\frac{L_{o}}{(1+t/\tau)^{2}}
\end{equation}
where $L_{o}=5 \times 10^{38}$erg/s, and $\tau=600$ years; no magnetic field is initially present in the SNR or in the ISM.
In all the simulations we have kept constant in time the following ratios of injected quantities: the ratio of  magnetic energy and total energy ($\sigma$), and that of particle energy and mass $\sim 100$. Units where $c=1$ are used.
Three different simulations will be considered here:
\begin{itemize}
\item Purely hydrodynamic; mass and momentum-energy injected as a wind with Lorentz factor 100, and $p/\rho=0.01$, no magnetic field.
\item Slightly magnetized wind ($\sigma \sim 0.003$); as in the previous case mass, momentum and particle energy are  injected as a wind with Lorentz factor 100, and $p/\rho =0.01$.
\item Magnetically dominated; in this case there is no self consistent quasi-steady shock solution on the timescale of PWN-SNR evolution (as soon as a shock is formed in the PWN, it starts collapsing back to the center). We have injected in the first cell magnetic and thermal energy with the ratio between the two close to equipartition (corresponding to $\sigma \sim 0.5$). In this case we have a hot source instead of a cold wind, with $p/\rho\sim 100.$
\end{itemize}

The injection of the wind is actually a delicate process, especially in the highly magnetized case. The reason  is that there are cases when the flow in the first cell becomes subsonic, due to the collapse to the center of the termination shock (this happens at very early times when a highly magnetized wind is considered, but it also happens in the low and zero magnetization cases, although at later times, since the decreasing pulsar input leads to situations in which the shock has to move very close to the pulsar in order for the wind ram pressure to contrast the inward push from the outer nebula). When the flow in the first computational cell is subsonic, there is no longer a complete freedom in the choice of the injected quantities. We have chosen to assign the values of mass, total energy fluxes and $\sigma$. This forced us to treat the magnetic field not as a primitive variable but rather as a derived quantity. This is especially important in the magnetic case to ensure energy conservation. Assigning the total energy flux and $\sigma$ fixes the magnetic luminosity $\dot E_{mag}$, so that:
\begin{equation}{}
E_{mag}(t+dt)=E_{mag}(t)+dt\times(\dot E_{mag}),
\end{equation}
and the value of $\dot B$ is then derived using the following equality:
\begin{equation}{}
E_{mag}(t+dt)=0.5\times(B(t)+dt\times\dot B)^{2}.
\end{equation}
We have checked that such injection condition ensures the correct energy balance in the PWN.

\section{Discussion}

In this section we will briefly review the various phases of the PWN-SNR interaction (the reader is referred to SW and BCF for a more detailed analysis in the hydrodynamic case), focusing on the effect of the magnetic field, especially inside the PWN, which had not been taken into account previously.

First of all, we find an overall slower evolution of the SNR, and, consequently, a slower evolution of the PWN, with respect to the work by SW, even if all the structures appear to be basically the same. As far as  the PWN is concerned, the slower evolution is mainly the effect of a different adiabatic coefficient (4/3 instead of 5/3) that makes the nebula more compressible (the PWN size is about 10\% larger if 5/3 is used, given the same energy input). As for the SNR, the different time scale of the evolution is most probably due to the different setup of the initial conditions (the energy released by the explosion is now in the form of kinetic energy rather than thermal). 

We do not expect the global evolution of the system to be different between the HD and RMHD case. The only two parameters that rule the evolution for a given SNR, are the PWN energy and its radius. In SW it is shown that, for a given SNR, the evolution of the contact discontinuity ($R_{pwn}$) is completely determined by the local value of the PWN pressure. When the RMHD case is considered, the total pressure may vary inside the PWN due to magnetic tension, but its value at the boundary still stays the same since it is a function of $E_{tot}$ alone. Neglecting adiabatic and radiation losses, the following evolution equation holds for the PWN in the general RMHD case:
\begin{equation}
\dot R_{pwn}^{2}= f(P(R_{pwn})) = f\left(\frac{E_{tot}}{4\pi R_{pwn}^{3}}\right), \label{Rpwn}
\end {equation}
where the function $f$ contains the dependences on the SNR parameters.

 Let us consider the two extreme cases: no magnetic field and
 magnetic field alone. When no magnetic field is present, the total pressure is just the thermal pressure and is approximately constant ($P_{tot}=P(R_{pwn})$) so that:
\begin{equation}
E_{tot}=\int_{0}^{R_{pwn}} 12\pi P_{tot}\ r^{2}dr= 4\pi P(R_{pwn}) R_{pwn}^3.
\end{equation}
On the other hand, in the magnetically dominated case $P_{tot}=B^2(r)/2$, and  $B(r)\sim 1/r$, so:
\begin{equation}
E_{tot}=\int_{0}^{R_{pwn}} 4\pi P_{tot}\ r^{2}dr= 4\pi P(R_{pwn}) R_{pwn}^3,
\end{equation}
and  the second equality in Eq.~(\ref{Rpwn}) still holds.

The thermal and magnetic pressure appear in a completely analogous form in the two cases (this is due to the fact that the different proportionality coefficient between pressure and energy for an HD (1/3) and magnetically dominated (1) plasma, exactly compensates the different pressure profile). So it is apparent that what is important for determining the propagation speed of the contact discontinuity is just the value of the total pressure, not the role that is separately played by the magnetic field and the particles. It can be also shown that the relativistic plasma and the toroidal magnetic field undergo the same adiabatic losses, so the energy variation is analogous for the two components under compression or expansion of the nebula (\cite{pacini}). As far as radiation losses are concerned, these are dynamically important only in the pressure dominated portion of the PWN ({\it i.e.} only for the time a particle spends in the pressure dominated region), and not in the magnetically dominated part where the dynamics is ruled by the magnetic field. Actually, radiative cooling may play a more important role in the evolution of the SNR. 

\begin{figure}
\resizebox{\hsize}{!}{\includegraphics{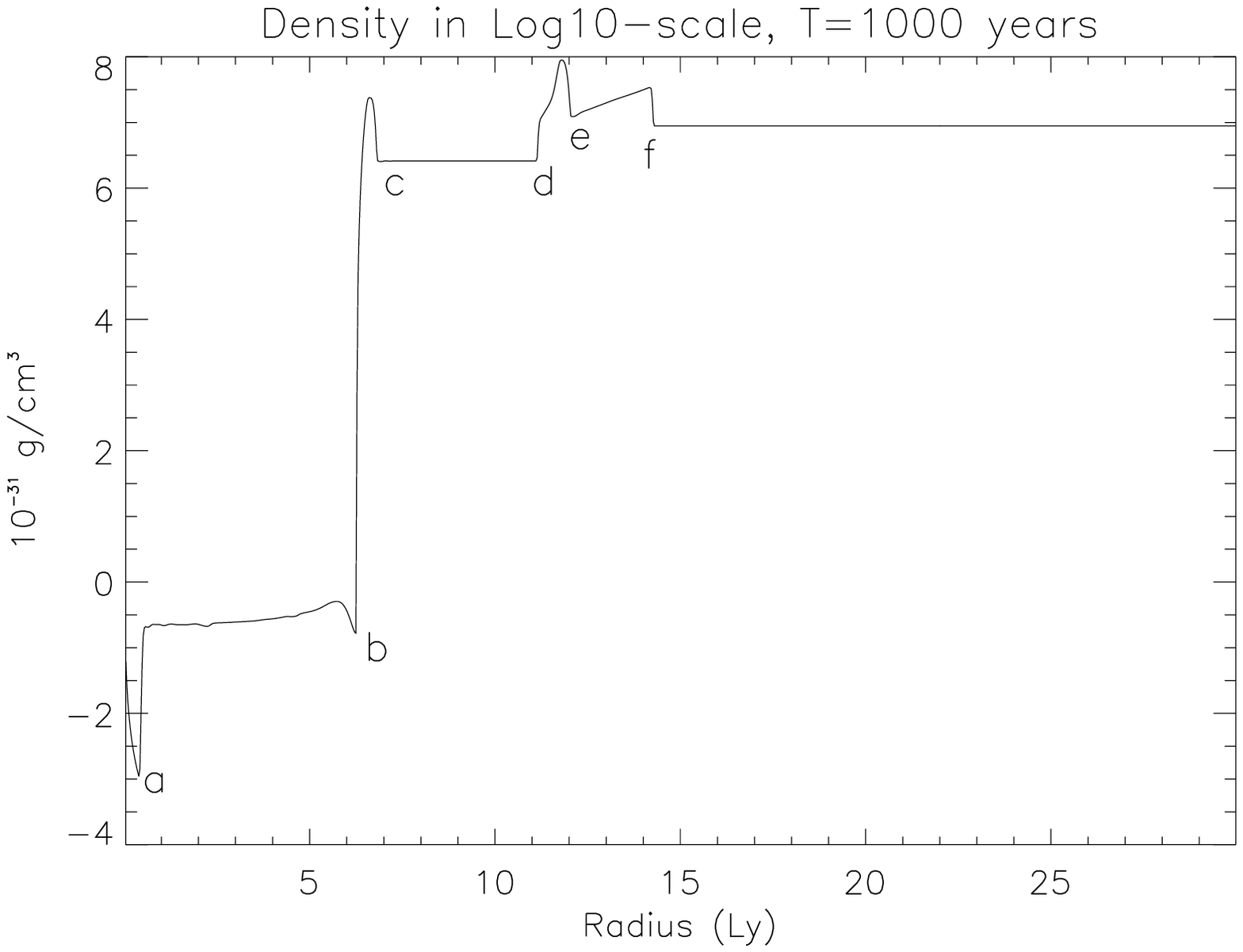}}
\resizebox{\hsize}{!}{\includegraphics{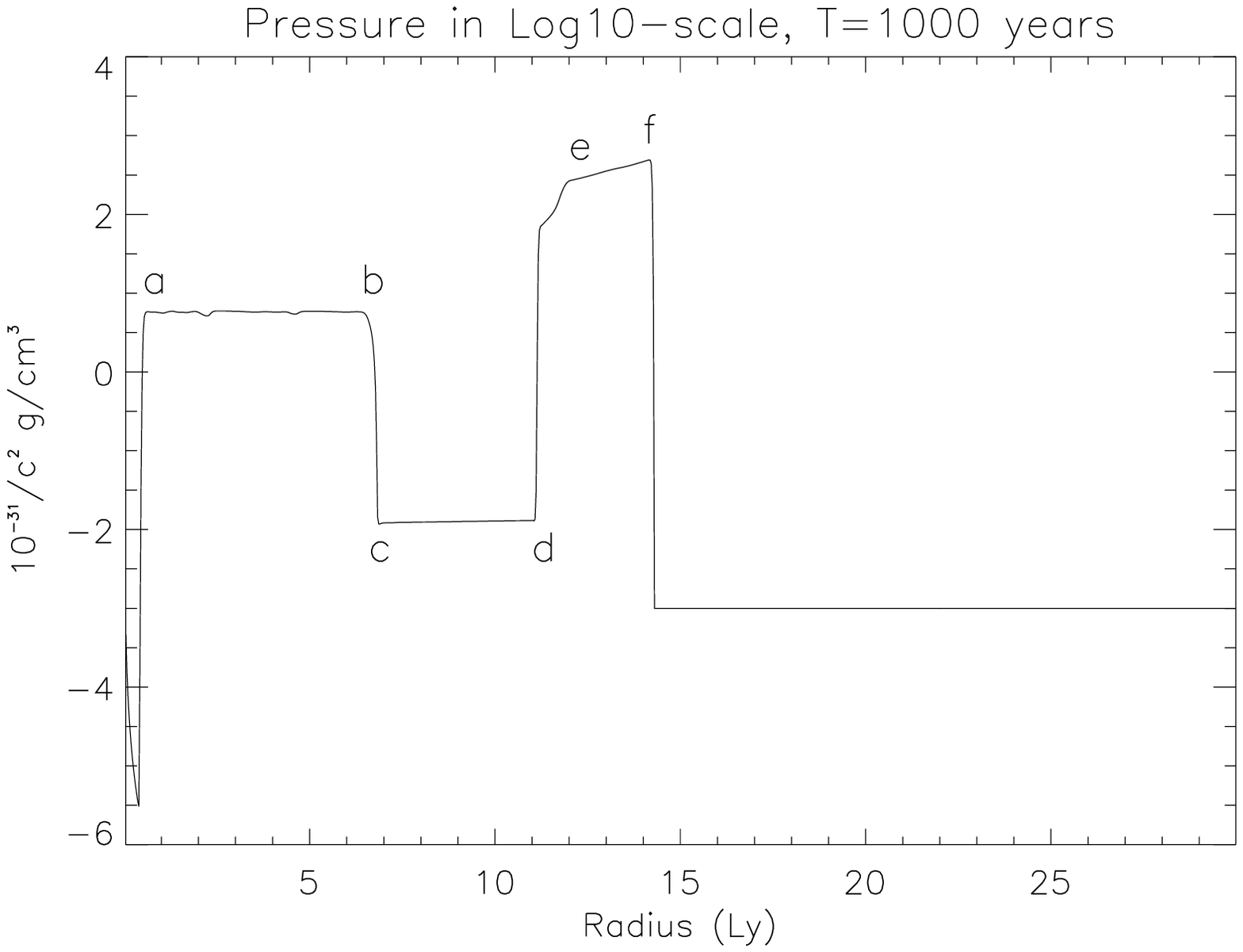}}
\caption{Density and pressure profile for the hydrodynamic case at time T=1 kyr after the supernova explosion. From the origin one can see the termination shock of the relativistic wind (a), the PWN contact discontinuity (b), the swept up shell of ejecta (c), the reverse shock (d), the contact discontinuity between ejected and shocked ISM material (e) and the blast wave propagating in the unperturbed ISM (f).}
\label{fig_hd}
\end{figure}

Fig.~\ref{fig_hd} shows the density and pressure after 1000 years in
the HD simulations: the small oscillations seen in the pressure and
density profile of the PWN, as well as the wall heating (point b) at the boundary (\cite{delzanna02}), are mainly due to the use of two different adiabatic coefficients (for a discussion of spurious  wave generation in multifluid simulations see \cite{kun98}), and they completely disappear when a single fluid is used. The various discontinuities are all well developed: moving from the origin outward, we find the pulsar wind termination shock, the contact discontinuity separating the hot relativistic bubble from the ejecta material, the shock in the ejecta driven by the PWN expansion, the reverse shock propagating into the ejecta, the contact discontinuity between the ejected and ISM material, and the blast wave propagating in the unperturbed ISM.

\begin{figure}
\resizebox{\hsize}{!}{\includegraphics{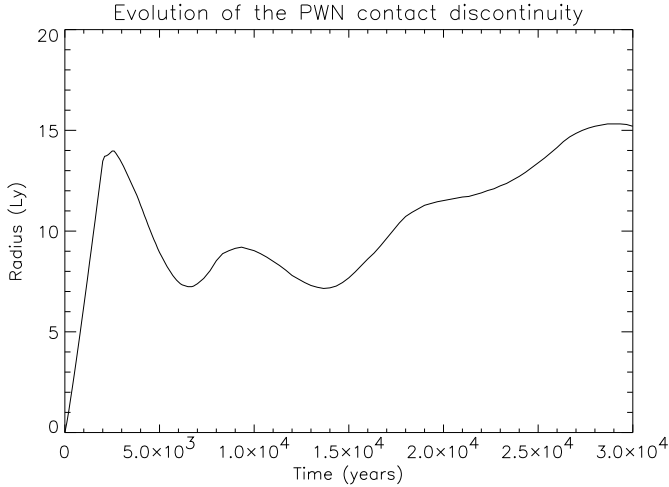}}
\caption{The radius of the PWN contact discontinuity as a function of time: T = 0-1.8 kyr free-expansion phase; T= 1.8-15 kyr unstable reverse shock reverberation phase; T= 15-30 kyr Sedov expansion phase. The curve shows the evolution for the hydrodynamic case, only differences too small to be shown on plot are present in the magnetic cases, mainly due to magnetic diffusion at the contact discontinuity.}
\label{fig_cd}
\end{figure}

In Fig.~\ref{fig_cd} the position of the contact discontinuity is shown for the HD case. The weakly and strongly magnetized cases are basically the same except for some minor differences mainly due to the numerical diffusion of the magnetic field. The only way to reduce this effect is to use more sophisticated wave-based Riemann solvers. However, we do not deem this necessary since the effect is kept within a few computational cells. As anticipated the behaviour is very similar to that found by SW, even if the evolution turns out to be slower. This agrees with what one would expect, given the same initial energetics. The free expansion phase lasts for about 2000 years, until the swept up shell of ejecta hits the reverse shock of the SNR. Then the expansion stops and the reverberation phase begins: the reverse shock compresses the PWN until the pressure of the latter becomes high enough to stop it and to push it back. There is a transient phase characterized by oscillations of the PWN contact discontinuity  and finally the structure relaxes to the Sedov phase. 

\begin{figure}
\resizebox{\hsize}{!}{\includegraphics{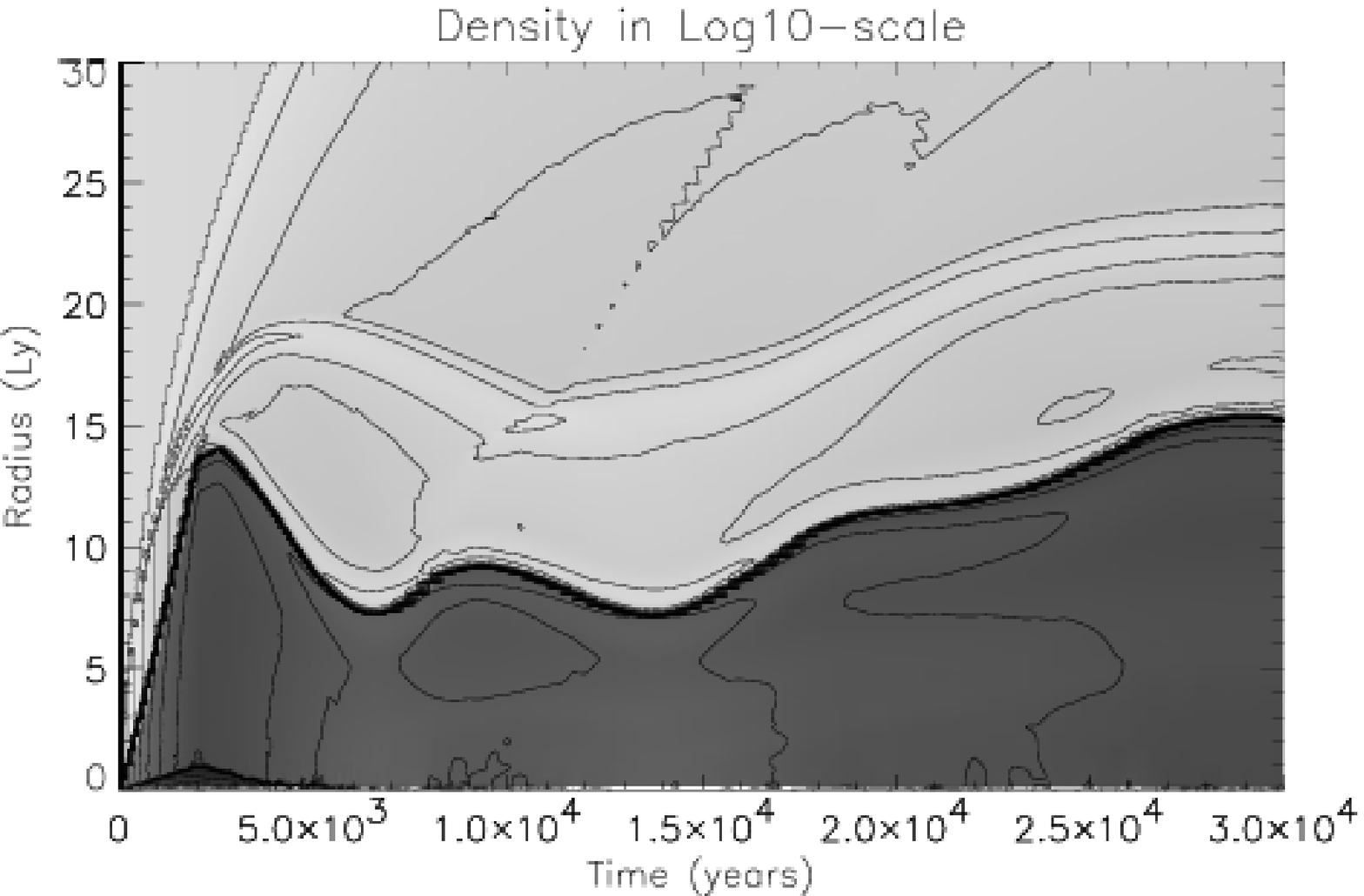}}
\resizebox{\hsize}{!}{\includegraphics{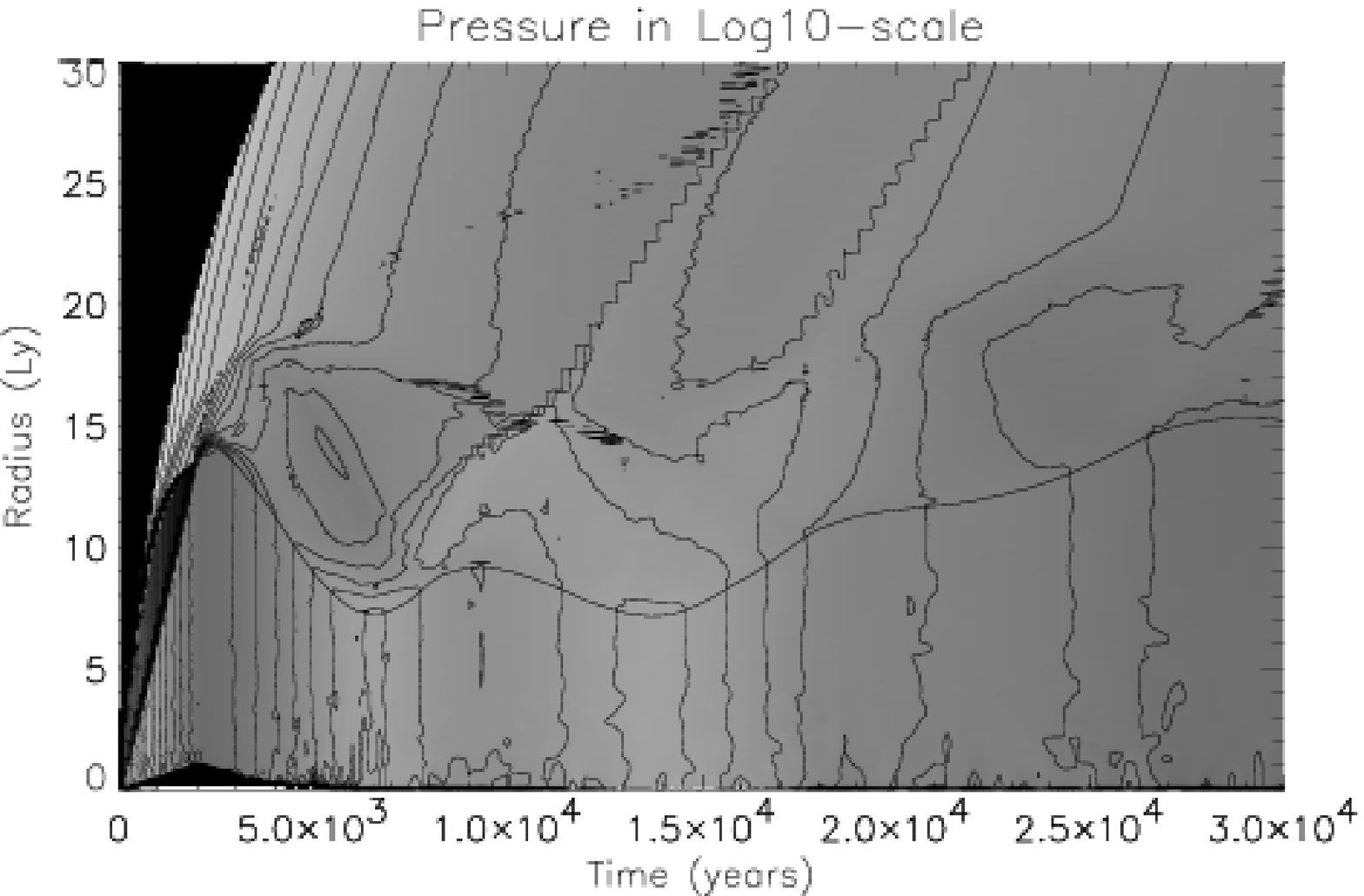}}
\caption{Evolution of a PWN inside a SNR for the HD case. Density (top) and pressure (bottom) in logarithmic gray-scale and contour levels, with black corresponding to low values and white to high values. In the bottom panel the position of the contact discontinuity is shown.}
\label{fig_ev1}
\end{figure}

 In Fig.~\ref{fig_ev1} the time evolution of both density and pressure is shown. As soon as the reverse shock starts compressing the PWN, the termination shock moves toward the origin and finally collapses to the first computational cell. At later times, when the SNR enters the Sedov expansion phase, the ram pressure in the wind has dropped to such a low value that the termination shock cannot be detached from the first cell any more. While the reverse shock moves inward compressing the PWN and increasing its pressure, the contact discontinuity separating the ejecta from the shocked ISM still moves outward. The pressure in the ejecta decreases due to rarefaction, and when the compression phase has gone on for about half of its total duration, it reaches the same values as the pressure inside  the PWN, and the compression starts slowing down. As can be seen from the pressure evolution, the PWN experienced phases of compression and rarefaction; if the latter are strong enough they may give rise to enhancements of the radio emission, so that one can expect to see old remnants with high radio luminosity near the pulsar position. These oscillations also trigger the formation and propagation of weak shocks in the SNR, which  may reheat  the ejecta. These shocks form when compressions stop and a new expansion phase begins (in our simulations we see only two of them, which propagate in the SNR with a velocity of the order of 1000 km/s, and a jump in pressure of about a factor of 2). As shown in BCF the number, amplitude and time scale for the oscillations during the reverberation phase, may vary in different PWN-SNR systems so that the effect of such weak shocks may be important in the case of low luminosity pulsars or light SN ejecta. 

In spite of the fact that the evolution of the contact discontinuity between the PWN and the ejecta (and as a consequence the evolution of the SNR) is the same in both the HD and RMHD cases, the structure inside the PWN shows different behaviors. This is due to the different stiffness of a magnetically dominated plasma with respect to a purely HD one. For a hot relativistic plasma such differences can be easily understood looking at the wave speeds: in the magnetically dominated  case the wave speed is $c$ while in the HD case it cannot exceed $c/\sqrt 3$, therefore the magnetically dominated region acts more similarly to a rigid body in transmitting inward information from the PWN boundary. This plays an important role during the reverberation phase. In a magnetized case the terminal velocity inside the PWN tends to a positive asymptotic value (that should match the velocity of the contact discontinuity), which increases as  the magnetic pressure becomes more and more important. When the PWN reaches the reverse shock in the SNR, the contact discontinuity cannot go any further and starts moving back to the origin. The magnetically dominated part (which is in asymptotic condition) also starts to move back and compress the particle dominated region, where mass and energy also increase as a consequence of the pulsar injection. A high density, and high pressure region is formed near the origin, whose dimension is a function of the ratio between the injected magnetic and thermal energy during the pulsar life (we have assumed it to be constant). In spite of this different behaviour, the magnetic field and particle pressure combine in a way to give the same total pressure at the boundary in the various cases, as we explained.

\begin{figure}
\resizebox{\hsize}{!}{\includegraphics{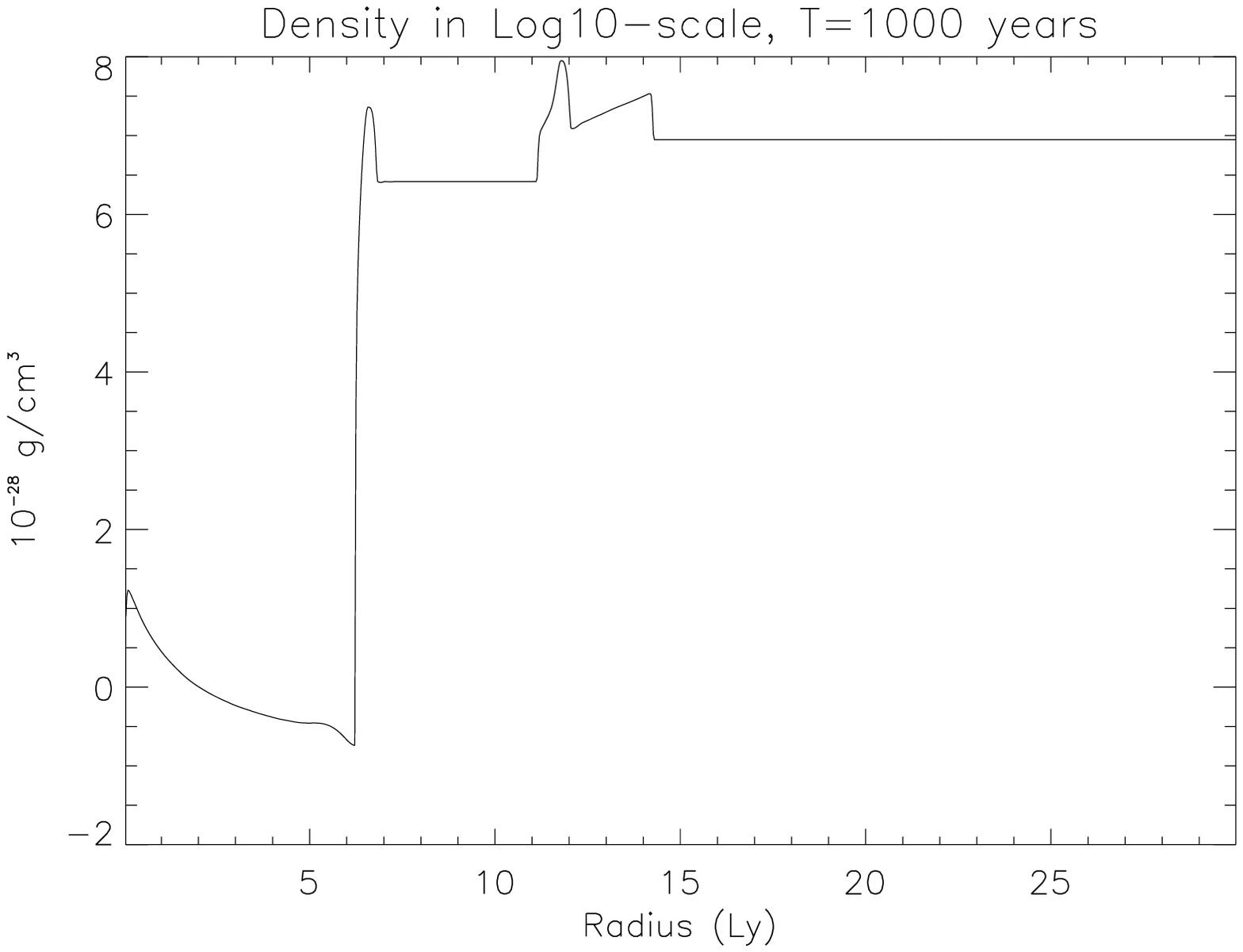}}
\resizebox{\hsize}{!}{\includegraphics{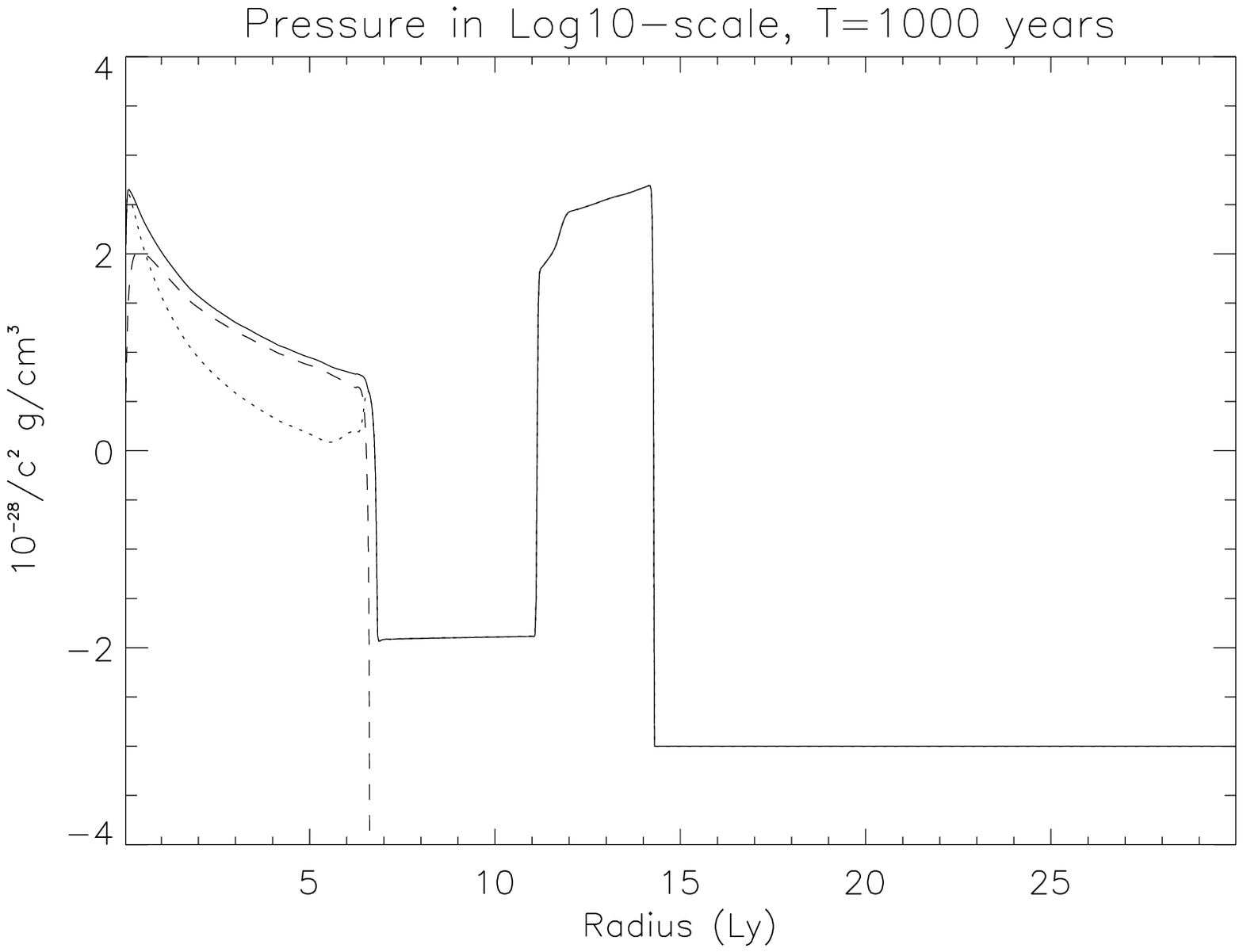}}
\caption{Density and pressure profile for the $\sigma = 0.003$ case at time T=1 kyr after the supernova explosion. The SNR structure is the same as in Fig.~2, while a different profile for PWN quantities arises: the termination shock is not detached; the total pressure at the PWN contact discontinuity is the same as in the HD case. The bottom panel shows the total (solid line), the thermal (dotted line) and the magnetic (dashed line) pressure.}
\label{fig_mhd1}
\end{figure}

\begin{figure}
\resizebox{\hsize}{!}{\includegraphics{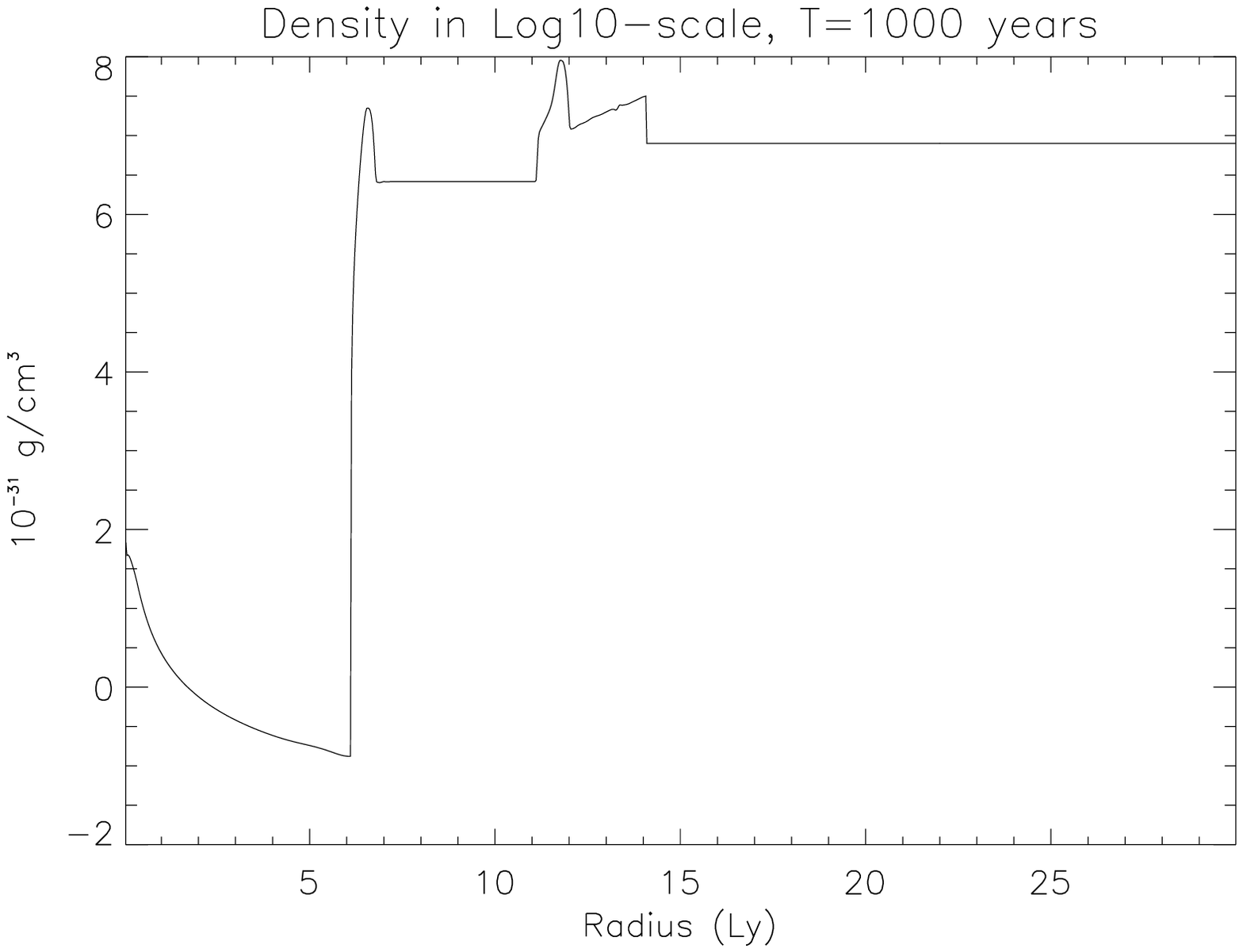}}
\resizebox{\hsize}{!}{\includegraphics{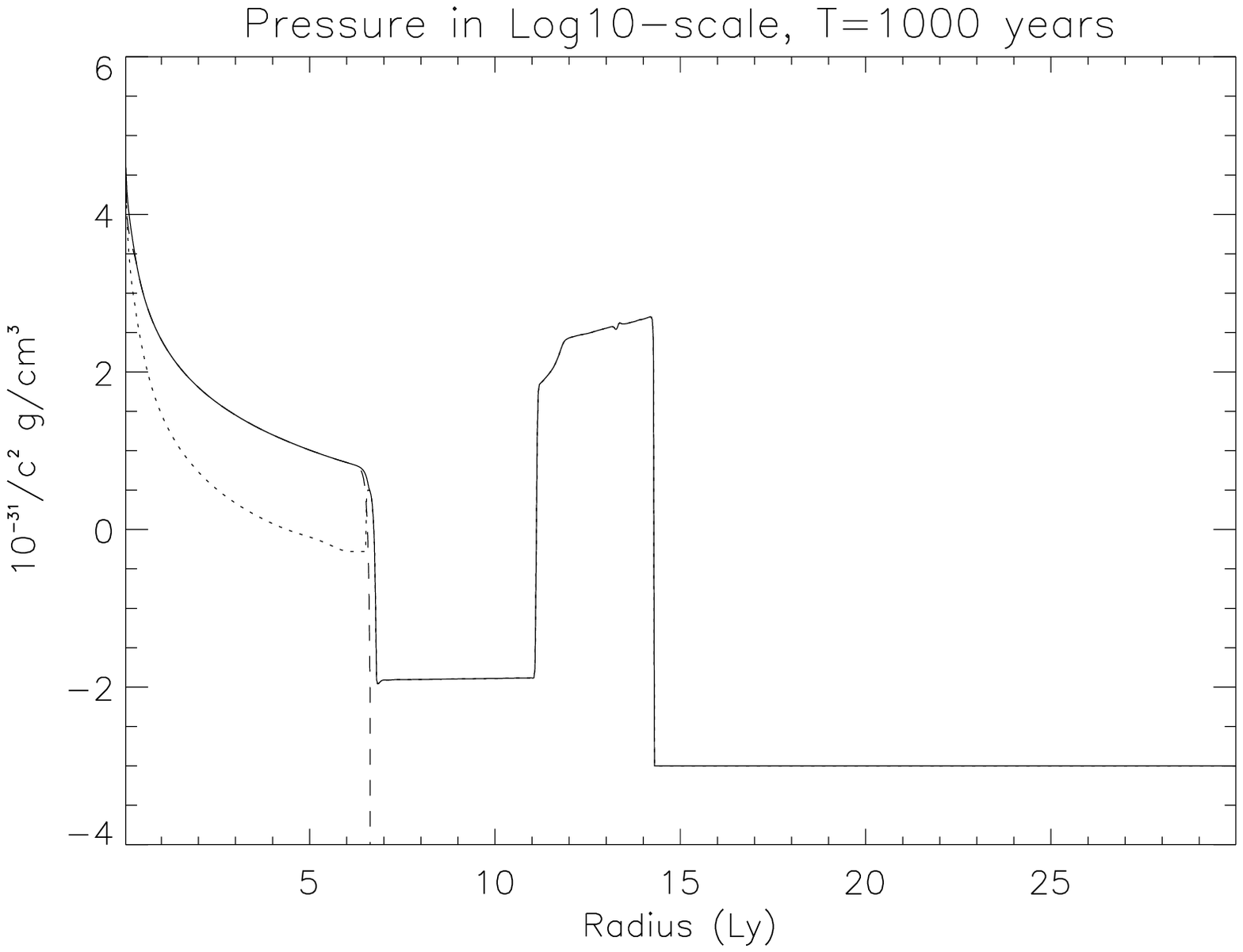}}
\caption{Density and pressure profile for the $\sigma \sim 0.5$ case at time T=1 kyr after the supernova explosion. In the bottom panel different curves represent various pressure contributions  as in Fig.~4}
\label{fig_mhd2}
\end{figure}

Fig.~\ref{fig_mhd1} and \ref{fig_mhd2} show density and pressure in
the slightly magnetized and magnetically dominated  case after 1000
years (the same as in Fig.~\ref{fig_hd}). As anticipated the position
of the contact discontinuity and the value of the pressure at the boundary are the same as the HD case, while rather different PWN structures arise.
The internal structure in Fig.~\ref{fig_mhd1} is consistent with the
Kennel \& Coroniti model (\cite{kennel84}) even if we are not able to
resolve the termination shock (this should be in the first cells but
numerical diffusion spreads it to the first one). The pressure profile
shows a central region dominated by the thermal pressure and an outer
magnetically dominated zone: at the contact discontinuity the PWN has
a ratio between magnetic and thermal pressure $\sim 10$. The wall
heating effect is still present even if it is less than in the HD
case. In Fig.~\ref{fig_mhd2} the magnetic pressure  is $\propto
r^{-2}$  (and almost coincident with the total pressure inside the
PWN) as expected for a magnetically dominated nebula (with a purely
toroidal magnetic field), only very near to the origin it reaches values close to equipartition with the thermal pressure.

\begin{figure}
\resizebox{\hsize}{!}{\includegraphics{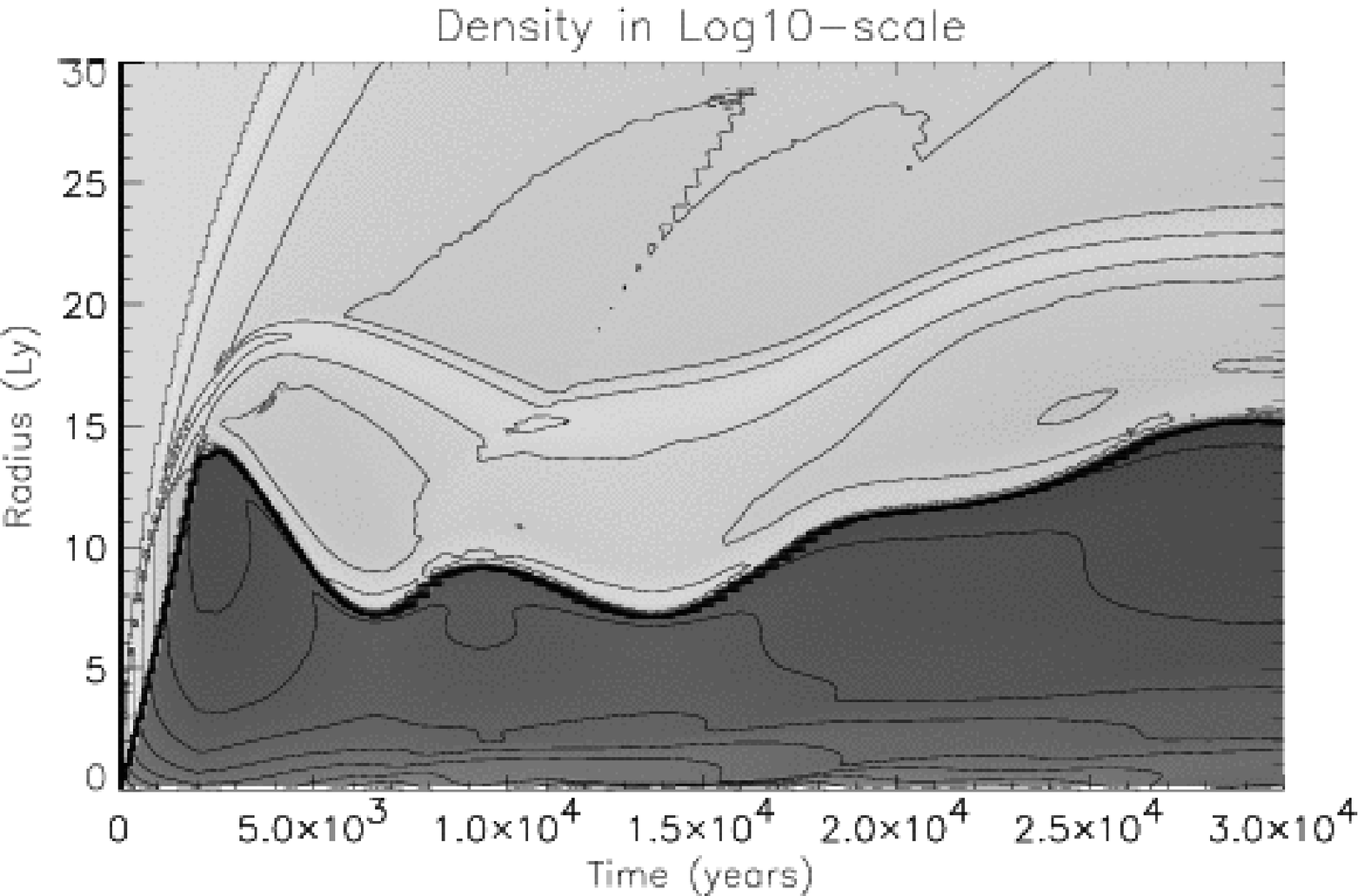}}
\resizebox{\hsize}{!}{\includegraphics{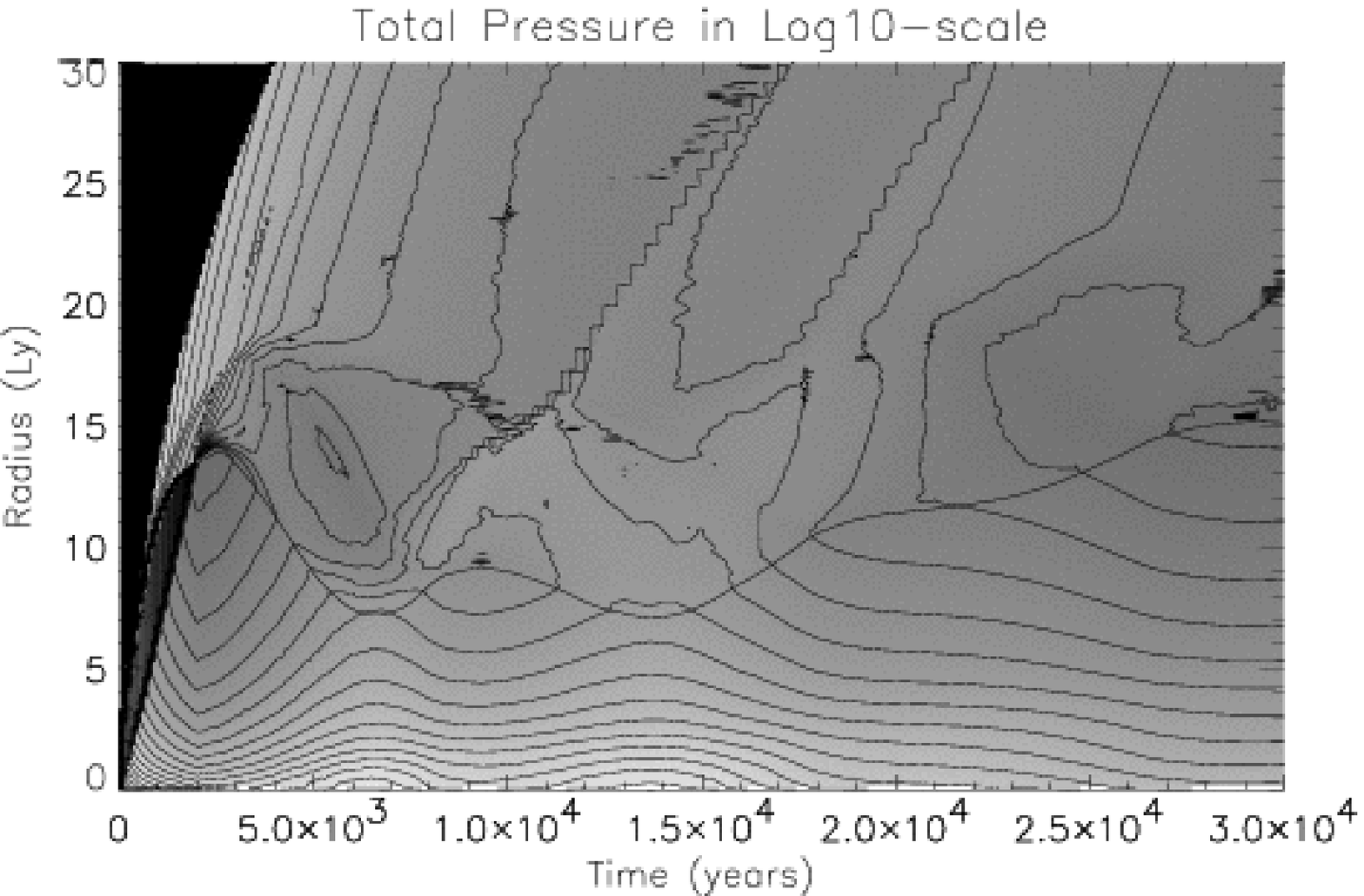}}
\resizebox{\hsize}{!}{\includegraphics{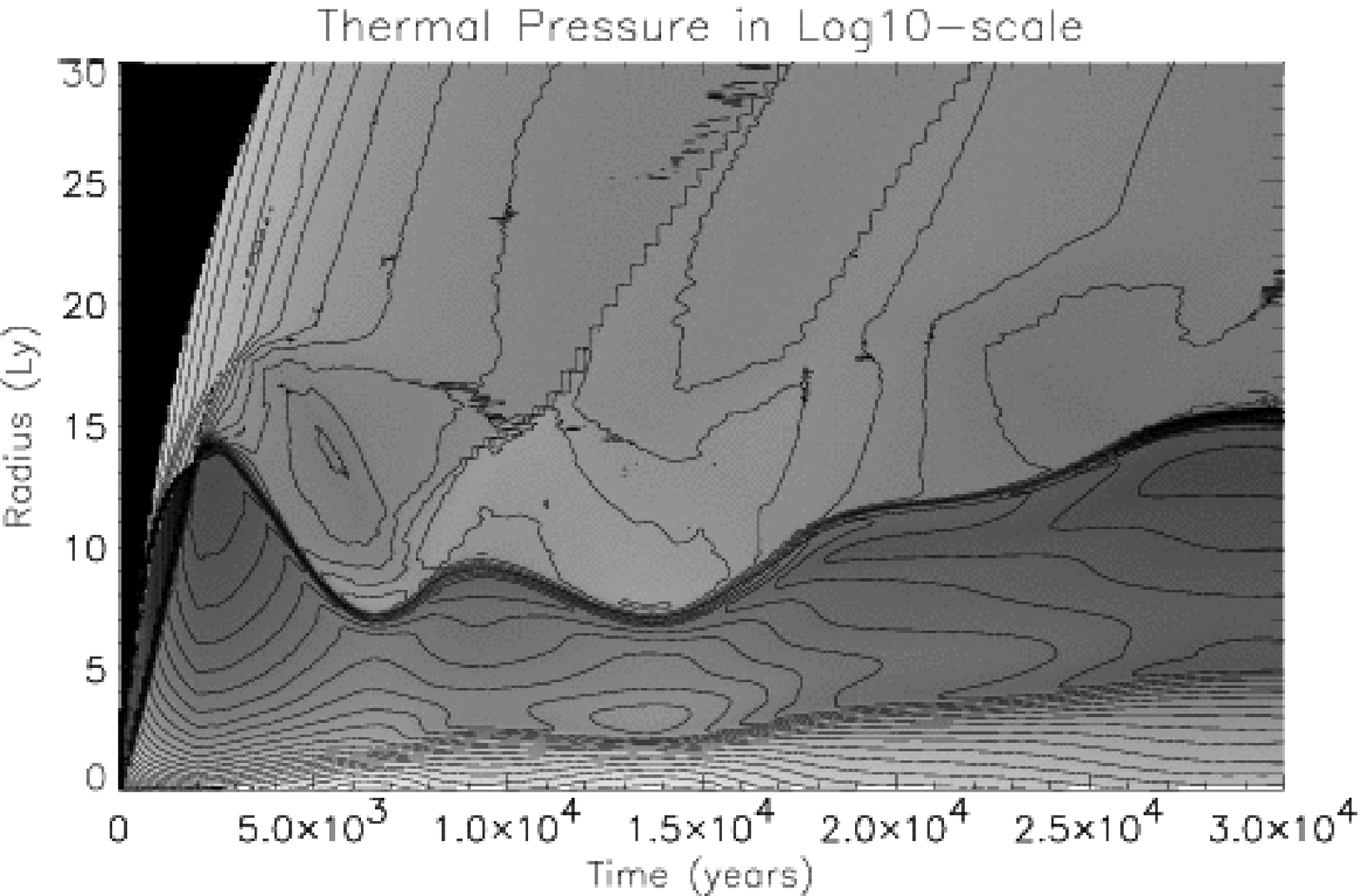}} 
\caption{Evolution of a PWN inside a SNR for the $\sigma = 0.003$ case. Density (top) and total pressure (middle) and thermal pressure (bottom) are represented in logarithmic gray-scale and contour levels, with black corresponding to low values and white to high values. In the middle and bottom  panels the position of the contact discontinuity is shown.}
\label{fig_ev2}
\end{figure}

In Fig.~\ref{fig_ev2} the temporal behaviour of density, total
pressure and thermal pressure is plotted for the case when $\sigma =
0.003 $: density and pressure in the SNR  are very similar to the
hydrodynamic case while inside the PWN they show different radial
profiles decreasing toward  the contact discontinuity. During the
various compression and expansion phases due to reverse shock
oscillations there are enhancements of pressure that can produce a
high temperature and high magnetic field region near the pulsar,
eventually observable at radio wavelengths. Looking at the thermal
pressure we can see the effect of the different rigidity between the
magnetically and thermally dominated part of the PWN. As compression
starts, the magnetic region moves suddenly and starts compressing the
thermally dominated portion near the origin. The material starts
moving toward the pulsar which still continues to inject mass and
energy so that these accumulate at small radii. Finally, when the
system enters the Sedov expansion phase, matter and thermal energy
start being advected away from the pulsar again. This behaviour is
completely absent in the HD case where density remains constant inside
the nebula. This fact is important in that it leads to expect
different observable signature in old SNR-PWN systems depending on the
magnetization of the bubble. The dimension of the overcompressed
region can in principle be used to estimate the magnetic field inside
the nebula and, from this, the magnetization in the pulsar wind
itself: the larger the size of the overcompressed region the lower the
magnetic field in the bubble. In the case of no magnetic field we find
a uniformly hot bubble, while in a magnetized case the emission
enhancement may be higher near the pulsar and, in principle, even easier to detect.

\begin{figure}
\resizebox{\hsize}{!}{\includegraphics{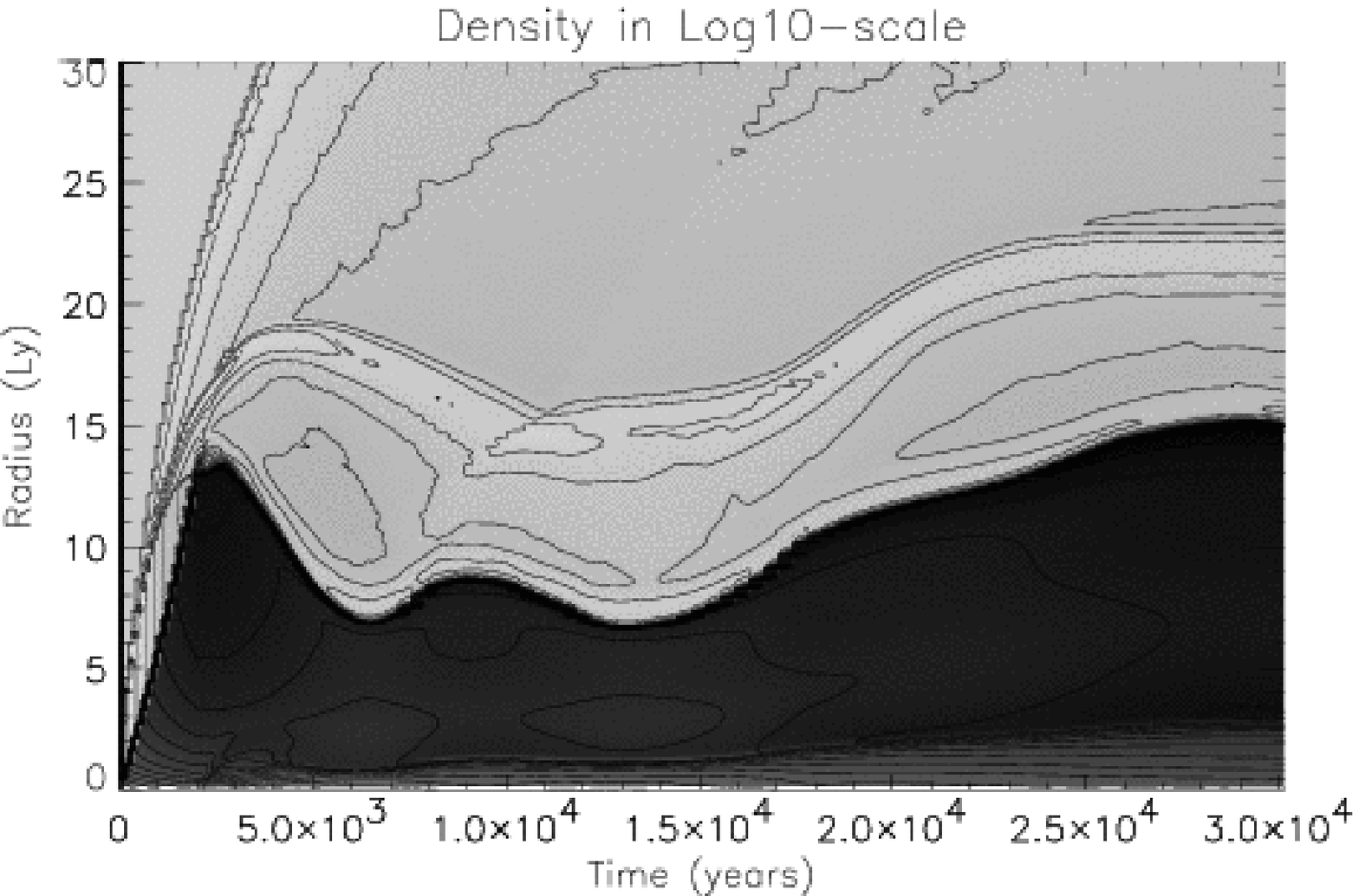}}
\resizebox{\hsize}{!}{\includegraphics{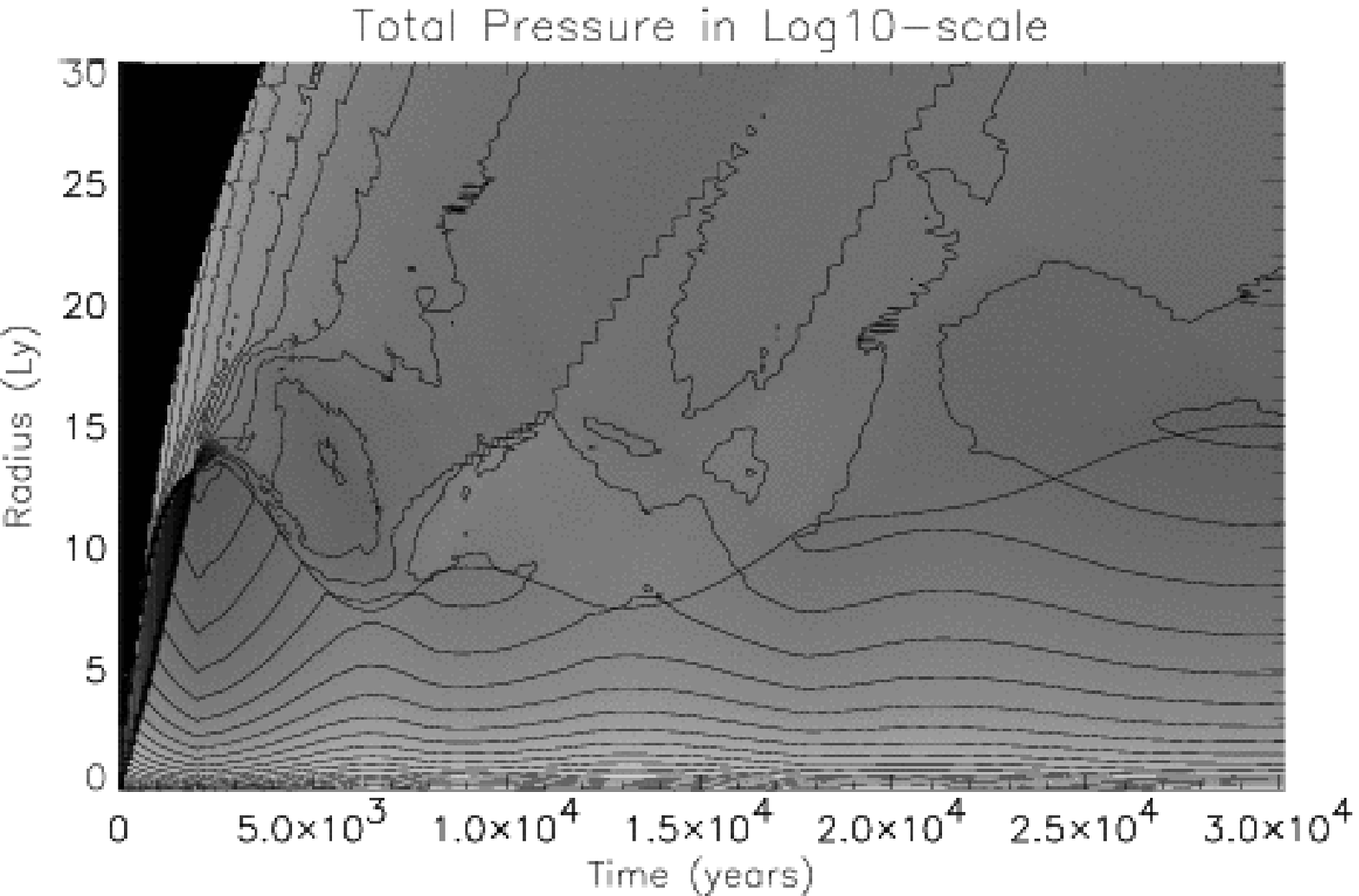}}
\caption{Evolution of a PWN inside a SNR for the $\sigma \sim 0.5$ case. Density (top) and pressure (bottom) are represented in logarithmic gray-scale and contour levels, with black corresponding to low values and white to high values. In the bottom panel the position of the contact discontinuity is shown.}\
\label{fig_ev3}
\end{figure}

A similar behaviour is present in the magnetically dominated case shown in Fig.~\ref{fig_ev3}, but in this case the thermal pressure becomes too noisy (it must be derived numerically from the total pressure, which is much larger, so that unavoidable accuracy errors arise), and the effect of the different rigidities is more easily appreciable by looking at the density evolution. The overall nebula can be considered to be in a magnetically dominated condition (except for the first cell). The injected material remains in the first cell during the compression and is advected away only at later times when the nebula enters the Sedov expansion phase. In this case we expect to find a high pressure and high magnetic field spot near the pulsar that may be hot enough to be detected not only in radio but also at higher frequencies (microwaves or IR).

\subsection{Comparison with existing analytic models}

Two classes of analytic solutions exist in the literature for the
internal structure of PWNs: the steady-state solution
(\cite{kennel84}, KC) and the self-similar solution (\cite{emmering87},
EC). It is interesting to compare the predictions of the analytic
models with the results of our simulations, also to evaluate how good  those
models are. We have chosen to compare our results with the EC model, since this allows for a non-zero velocity of the termination shock. 

It should be emphasized that the applicability of these models is limited to the first phase of the evolution of the PWN-SNR system, before the reverse shock in the SNR reaches the PWN contact discontinuity. Moreover a comparison between our results and the work by EC is only possible for the HD and slightly magnetized case, the only two case for which a shock solution exists.

The solution found by EC (as well as that by KC) relies on two strong assumptions: a constant
pulsar spin-down luminosity and a constant velocity at the outer
boundary of the PWN. None of these applies to our case: in our
simulations the spin-down power fed to the nebula by the PSR varies
according to Eq.(3) and the velocity of the contact discontinuity
increases with time during the evolutionary stage we are considering,
changing from an initial value of $0.005c$ to a value of $0.007c$ when the
system is 2000 years old. Both these factors lead to differences
between our results and those found by EC, but surely the different
energy input as a function of time plays the most important role. As
the PSR luminosity decreases, the position of the wind termination
shock moves to smaller radii than in the constant luminosity
case. This effect can be easily estimated in the pure HD
case. Assuming a constant energy input, the position of the termination
shock $R_{s}(t)$ can be estimated through the condition of pressure
equilibrium:
\begin{equation}
\frac{L}{4\pi c R_{s}^{2}(t)}=\frac{Lt}{4\pi R_{pwn}^{3}}\; ,
\end{equation}
this becomes in the time dependent case:
\begin{equation}
\frac{L(t)}{4\pi c R_{s}^{2}(t)}=\frac{\int_{0}^{t}L(t)dt}{4\pi R_{pwn}^{3}}.
\end{equation}
When the PSR luminosity is described by Eq.(3), after 1000 years, the
value of $R_{s}$  is a factor 0.612 less than
in a case when the luminosity is taken to be constant and such to give
the  the same value of $R_{pwn}$.

 Taking the EC solution in the unmagnetized case and with the shock
 velocity set to $3.5 \times 10^{-4}c$ (as evaluated from our
 simulation), we can determine the position of the contact
 discontinuity (normalized to the termination shock radius) as the point where
 the flow speed equals the values of $7. \times 10^{-3}c$, corresponding to the velocity of the contact
 discontinuity in  our
 simulation). Doing this, we find $(R_{pwn}/R_s)_{EC}\sim 0.5
 (R_{pwn}/R_s)_{sim}$, where the subscript $_{EC}$ is the ratio computed on the
 basis of EC model, and the subscript $_{sim}$ is our simulation value. This is consistent with our expectations based on the discussion above.

\begin{figure}
\resizebox{\hsize}{!}{\includegraphics{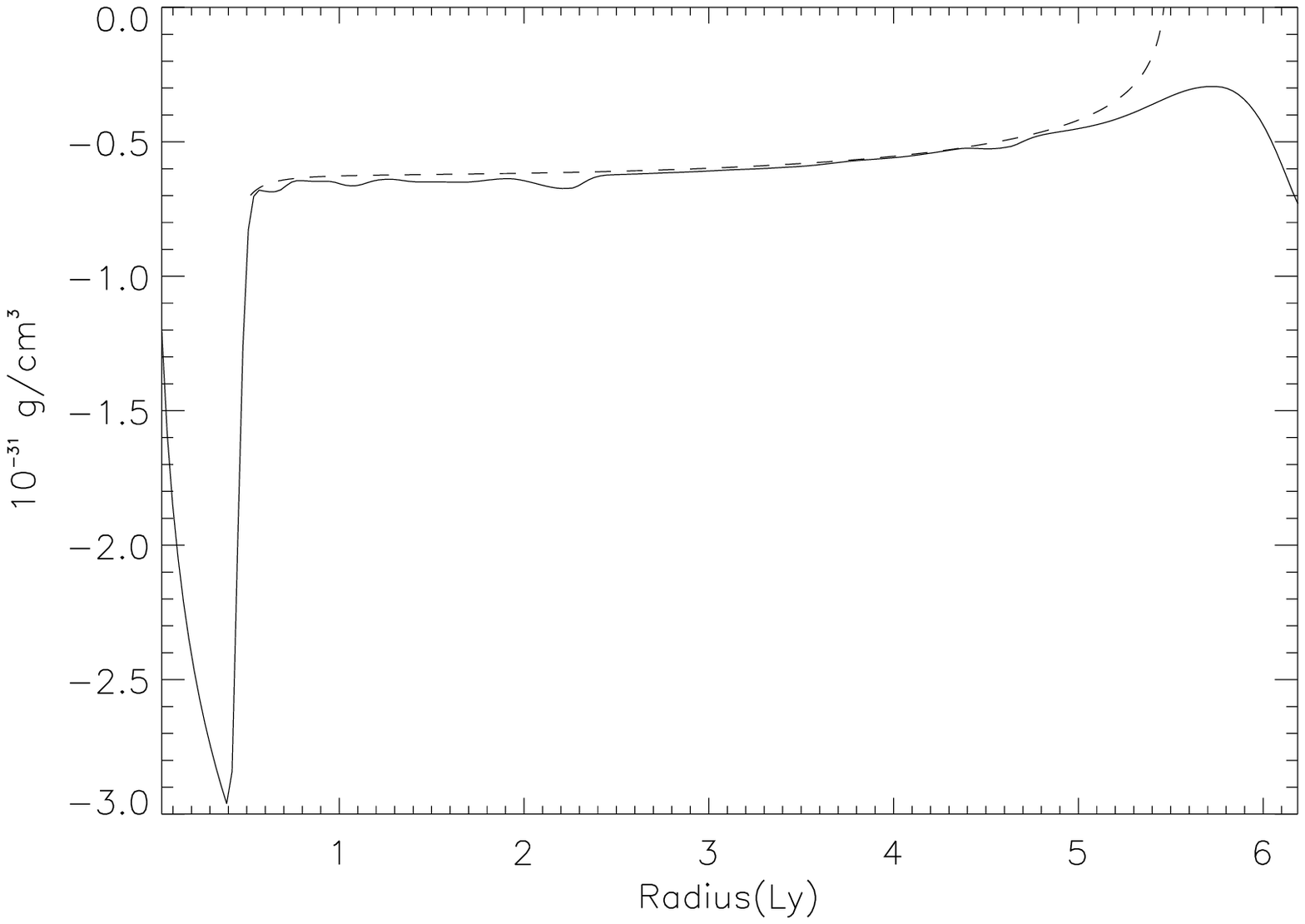}}
\resizebox{\hsize}{!}{\includegraphics{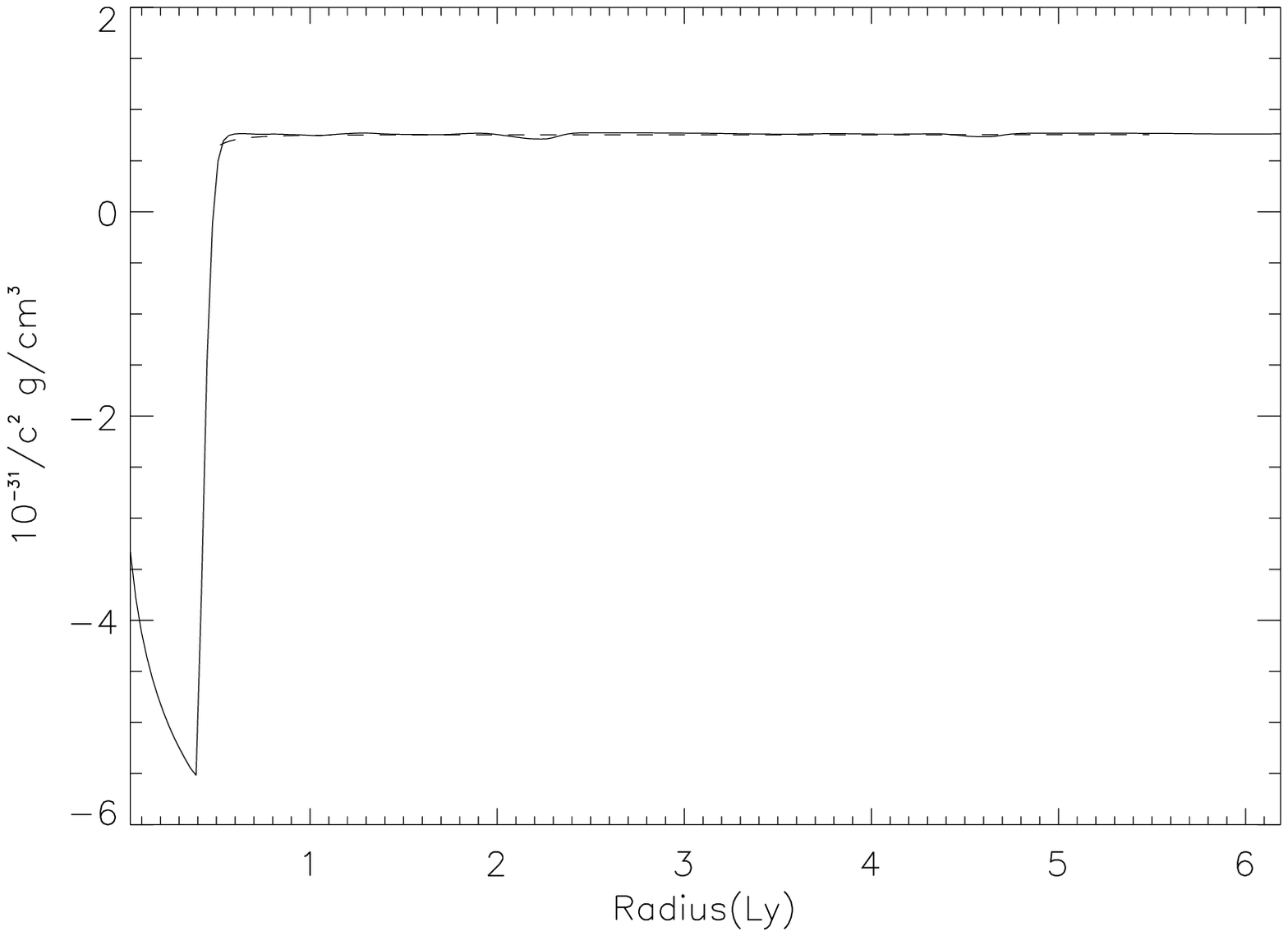}}
\caption{Comparison of the density (upper) and pressure (bottom) profiles in the pure hydrodynamic case. The solid line is our numerical simulation while the dashed line is the EM model rescaled to the termination shock radius.}\
\label{fig_conf1}
\end{figure}

In Fig.~\ref{fig_conf1} the numerical simulation is compared with the
EC solution scaled to the same wind termination shock radius. Apart
from the small waves previously discussed, we see a good agreement in
the post shock region. The two models show bigger differences in the
outer part of the nebula. As we mentioned, the EC model gives a
smaller nebular radius: the outer nebula is ``more stretched'' in our
case (we want to point out that even if the analytic solution exends
up to about 5.5 Ly, the dimension of the nebula determined by matching the
boundary velocity is smaller, $\sim$3.6 Ly). This difference is the result of the different pulsar spin-down law: material in the outer part was injected at early times, when the luminosity was larger, and carries more energy, so that it tends to expand as the wind ram pressure drops, and to push the more recently injected material close to the pulsar.

\begin{figure}
\resizebox{\hsize}{!}{\includegraphics{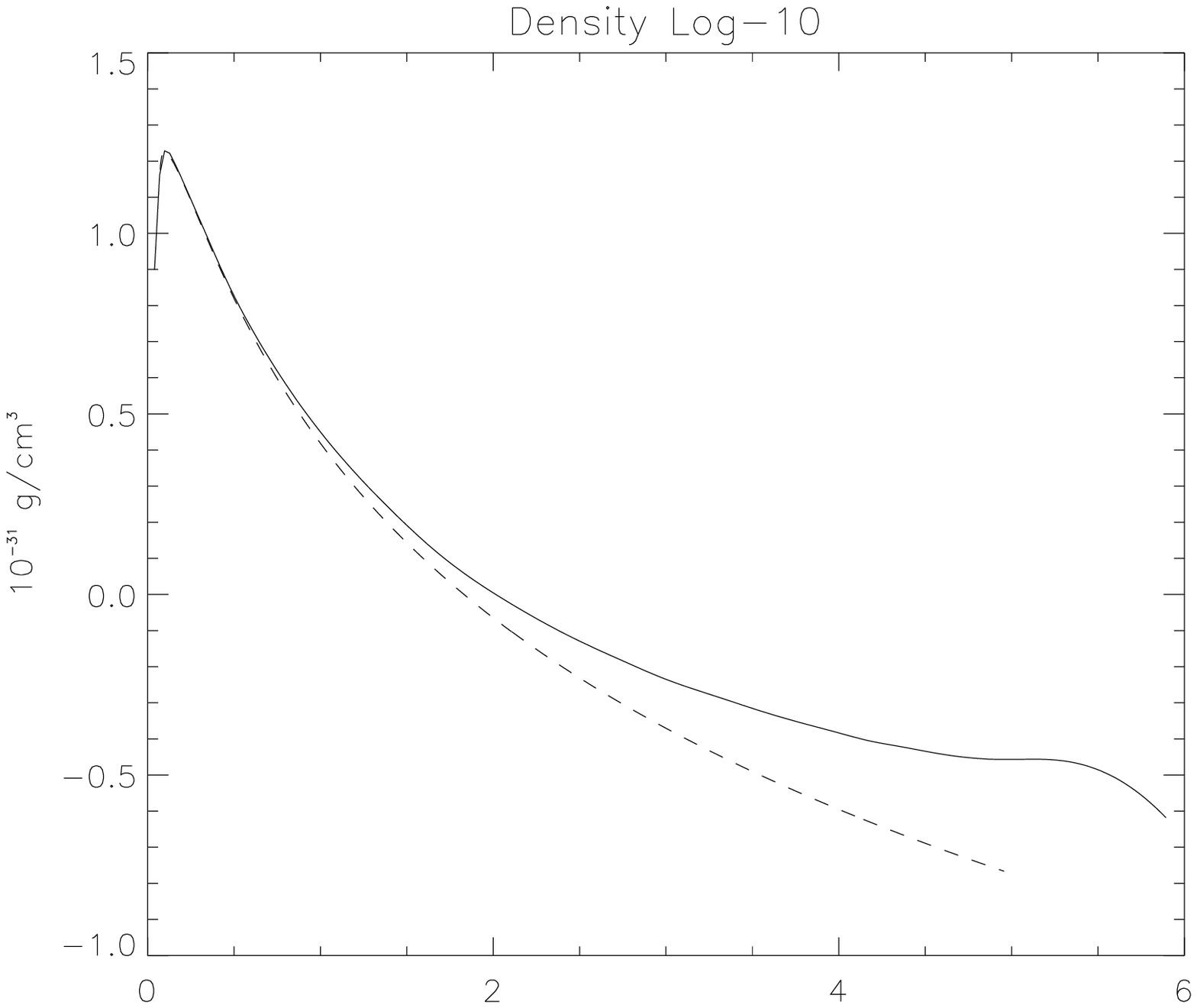}}
\resizebox{\hsize}{!}{\includegraphics{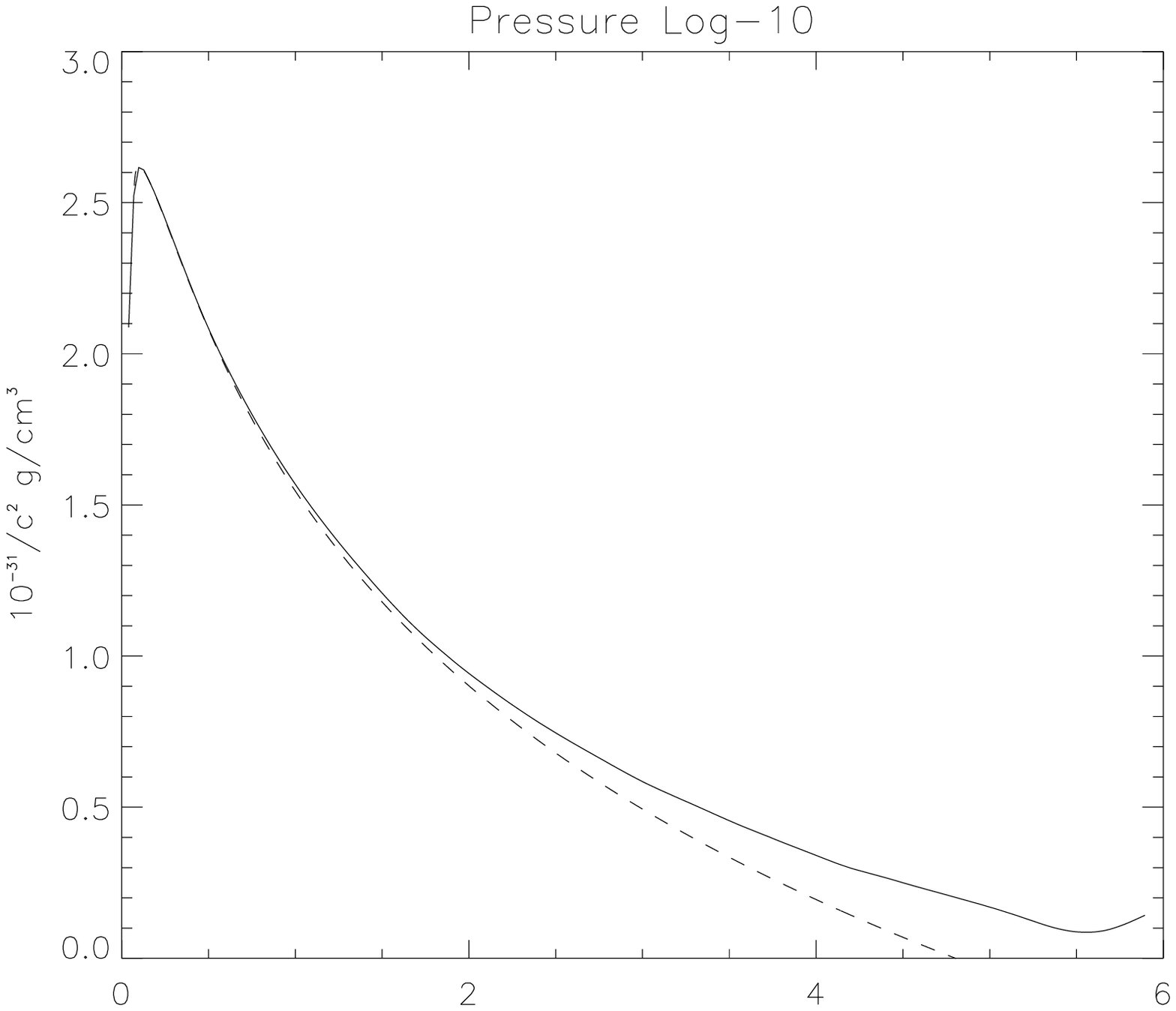}}
\resizebox{\hsize}{!}{\includegraphics{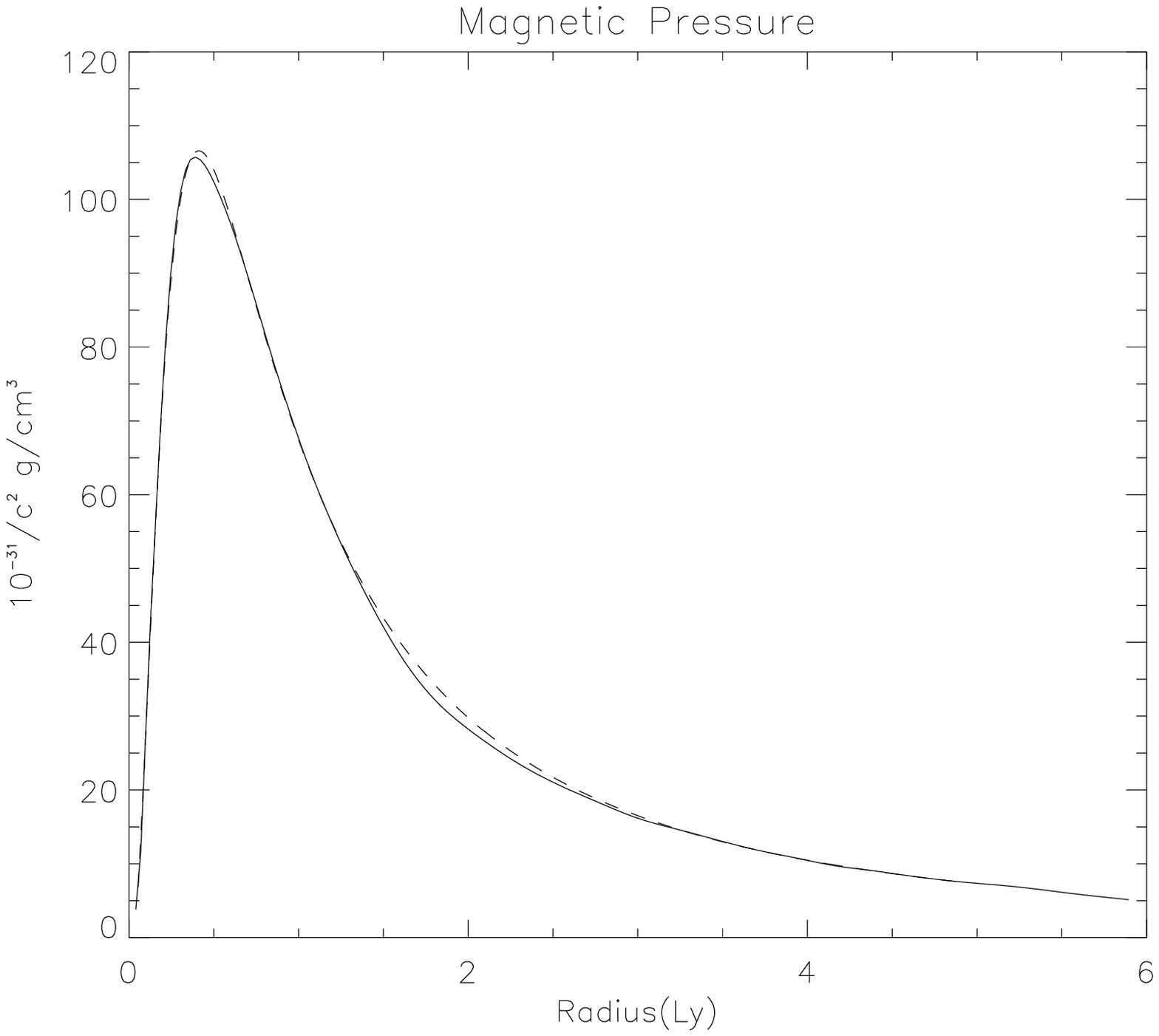}}
\caption{Comparison of the density (upper) and thermal pressure (middle) and magnetic pressure (bottom) profiles for the case $\sigma=0.003$. The solid line is our numerical simulation while the dashed line is the EM model rescaled to the termination shock radius.}\
\label{fig_conf2}
\end{figure}

\begin{figure}
\resizebox{9cm}{!}{\includegraphics{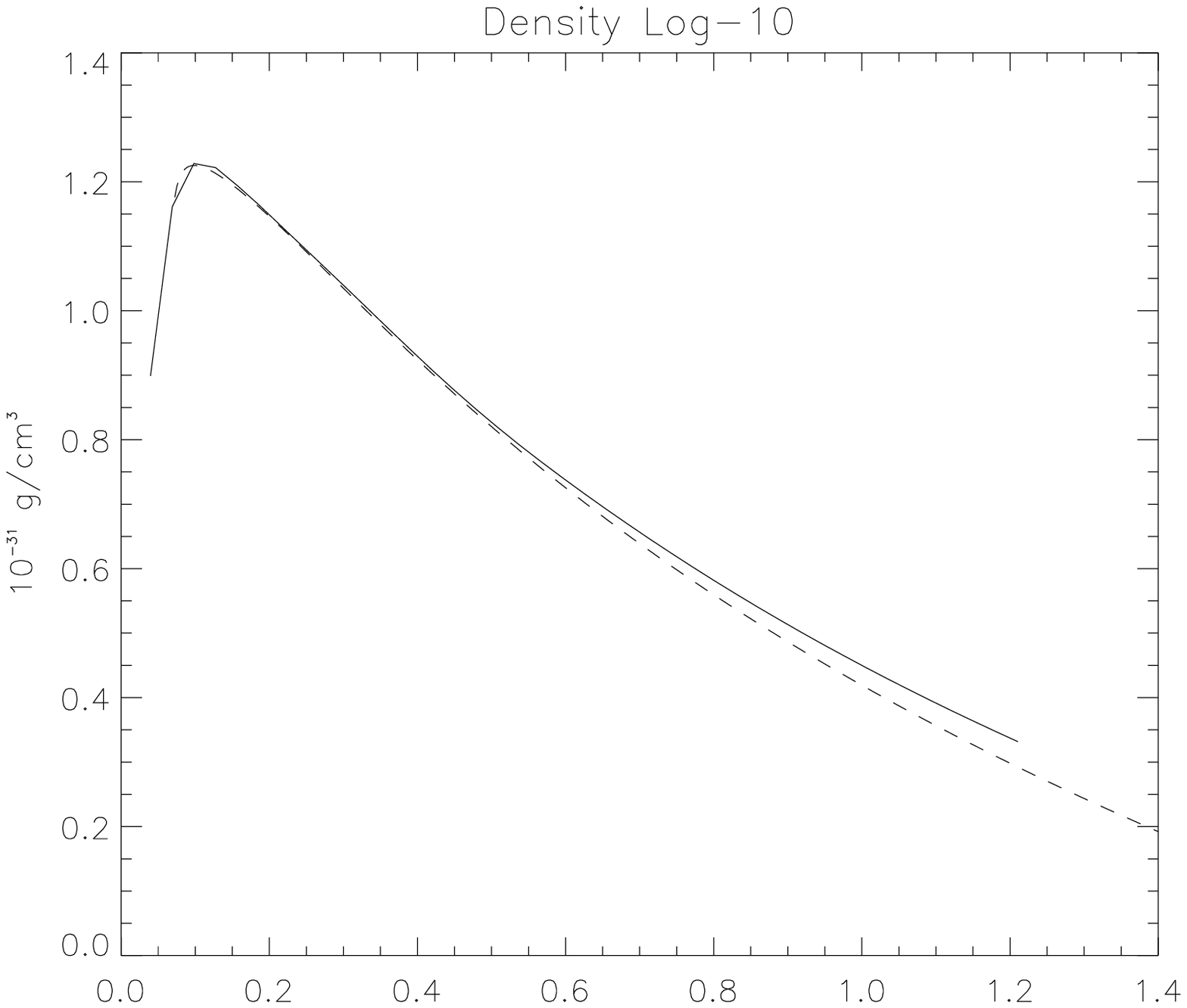}}
\resizebox{9cm}{!}{\includegraphics{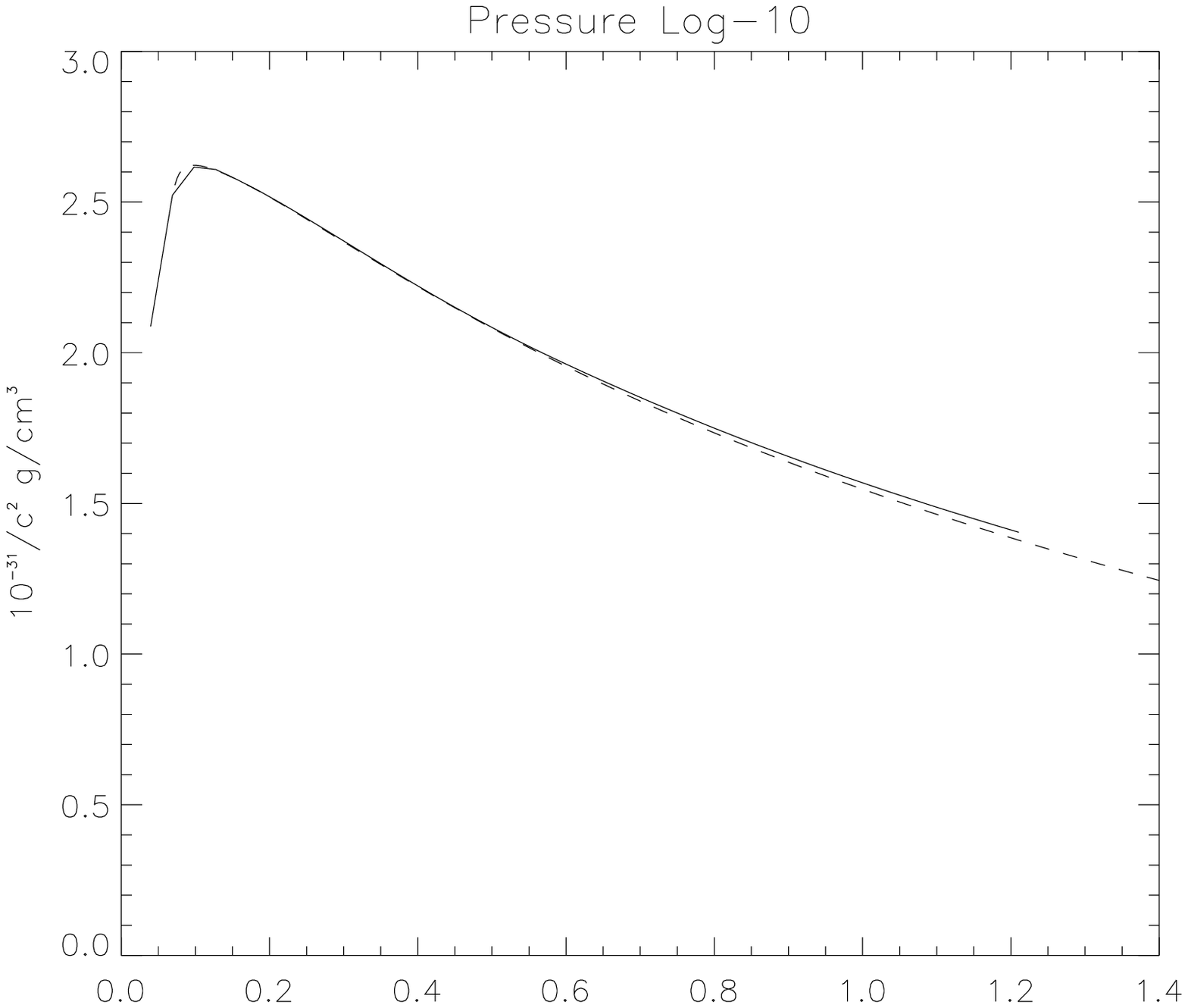}}
\resizebox{9cm}{!}{\includegraphics{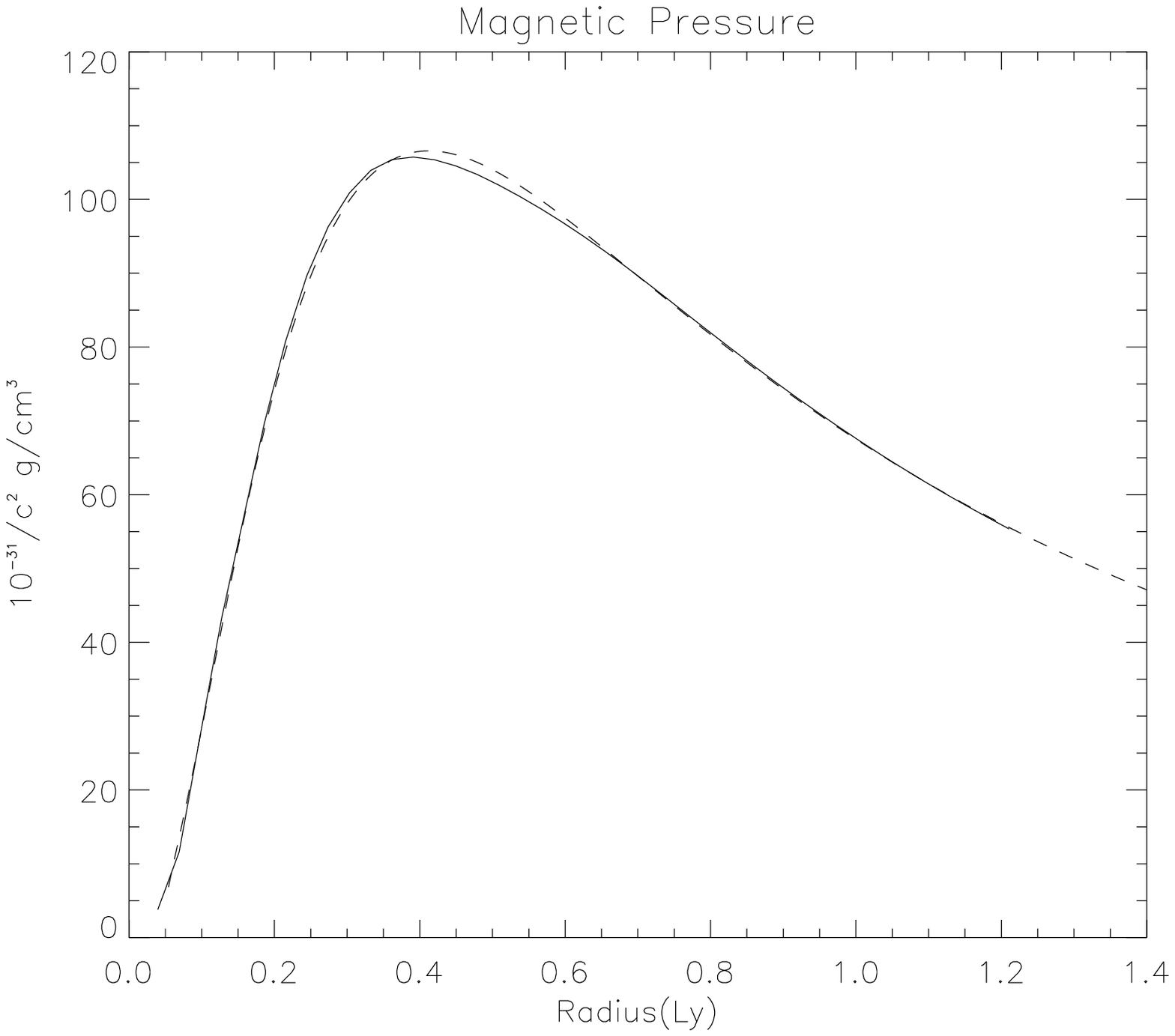}}
\caption{Same as Fig.10 but for the immediate post shock region.}\
\label{fig_conf3}
\end{figure}

A comparison in the $\sigma=0.003$ case is even more delicate: the variation of the boundary velocity is now more important because the velocity is close to the asymptotic value for a shocked wind with such magnetization so that even small variations can lead to major differences in the nebular size. Again we have chosen the EC model with the same magnetization for a comparison, even if in this case the termination shock is not detached from the first cell so that we could easily assume that its velocity is zero. 
Results are shown in Figs.~\ref{fig_conf2} and ~\ref{fig_conf3}, both
for the overall nebula and for the post shock region. The analytic
model has been rescaled to the shock position resulting from our simulation
(determined through a best fit procedure)
 which turns out to be at about 0.07 Ly from the
origin. The radial profiles agree very well with the EC model, even if
the nebula is more extended and the magnetically dominated part appears to
be larger. The differences in dimension can again be in part accounted
for as an effect of the different pulsar spin-down law ($R_{S}$ should
be 0.2 Ly if $L(t)$ were constant), but likely, in this case, variations of the
velocity at the boundary as well as numerical discretization are also of some importance.

Our results suggest that, taking into account the spin-down process, the Crab Nebula, where the ratio $R_{pwn}/R_{s}$ is believed to be about 20, should probably have a smaller magnetization parameter than what is estimated by KC. It should be noted that EC arrived to the same conclusion based on the speed of the termination shock. However this subject deserves further investigation.

\section{Conclusion}

In the present paper relativistic MHD simulations of the evolution of a PWN inside a SNR, in the spherically symmetric approximation, are shown for the first time. The early free expansion and Sedov-Taylor stages have been studied, both in the hydrodynamical and magnetohydrodynamical (with different magnetizations) regimes for the PWN. These simulations are mainly aimed at extending previous similar works (especially SW) to the more appropriate relativistic and magnetic regimes. Moreover, two different adiabatic coefficients are used for the PWN and for the SNR (two fluids).

An impoertant result is that the evolution of the contact discontinuity only depends on the total energy in the PWN while it is independent on its magnetization. The same result was previously found by \cite{emmering87} under the assumptions of time-constancy of the PSR power input and of the velocity of the contact discontinuity.

Our simulations do not take into account radiative losses. Such losses  may affect the PWN in the early phases when synchrotron losses are especially intense, but, in principle, on the time scale of the PWN-SNR system evolution they could be accounted for by renormalizing the pulsar spin-down energy input (and eventually changing the magnetization of the nebula). Radiative losses may be important in a non-trivial way for the dynamics of the contact discontinuity and the SNR evolution.  Radiation cooling may affect the dynamics of the swept up shell of ejecta as well as the pressure evolution during the reverberation phase eventually reducing the time-scale and the amplitude of the oscillations of the contact discontinuity. The main simplification of the model is, however, the reduction of the degrees of freedom of the system because of the assumption of spherical symmetry (i.e. 1-D). This prevents us from addressing the very important issue of how the instabilities, that are known to occur at the interface between the PWN and the ejecta, both during the expansion and reverberation phase, develop and evolve. At the same time we completely neglect the intrinsic anisotropies that the system is likely to posses, like those due to the pulsar spatial velocity or to the internal structure of the ejecta.

We also found that the time-dependence of the spin-down process of the pulsar and the decrease of ram pressure in the wind might be an important element in the time dependent modelling of PWNs. Steady-state models can provide a good description of the radial dependence of the various quantities (in units of the distance from the shock) but they cannot be used to estimate with confidence the relative dimension of the contact discontinuity and the wind termination shock. This argument however deserves further investigation which is beyond the scope of this article.

As in previous studies we find that the transition between the free expansion phase and the Sedov-Taylor expansion are unsteady and characterized by reverse shock reverberation that increases both the pressure and the magnetic field in the PWN, eventually reheating the plasma and producing bursts of radio emission in the later phase of the PWN life, when the pulsar has radiated almost all its spin-down energy. We do not find significant differences in the magnetic case. However, magnetic forces inside the nebula give pressure and density profiles with higher values near the origin (provided the total energy is the same). Moreover, the pressure at the boundary keeps the same value and the evolution of the contact discontinuity and of the SNR is unchanged. Even compression does not alter the global behaviour of the system, but the presence inside the PWN of two different regions (an internal thermal pressure dominated and an outer magnetic dominated zone) with  different associated rigidities gives rise to an overcompression near the origin, where mass and energy coming from the pulsar are confined inside a small radius. In this case a high-temperature and high-magnetic field spot is formed, which could eventually be revealed not only in radio but also at higher frequencies (microwaves or IR).

\begin{acknowledgements}
The authors thank  Rino Bandiera for the fruitful discussion and the
reading of this article, and referee R. Chevalier for his useful
comments and especially for asking us to include a comparison with
analytic models. This work has been partly supported by the Italian Ministry for University and Research (MIUR) under grants Cofin2001-02-10, and partly by a SciDAC grant from the US Department of Energy High Energy and Nuclear Physics Program. We thank the North Carolina Supercomputing Center for their generous support of computing resources.

\end{acknowledgements}


\end{document}